\documentclass[11pt,a4paper]{article}
\usepackage{jheppub}
\usepackage[T1]{fontenc}
\usepackage{bm}
\usepackage{booktabs}

\title{\bf dSIBP: a package for generating IBP seeds for QFT in de Sitter spacetime}

\author[a,b]{Jiaqi Chen}
\author[c]{Yi-Xiao Tao}
\affiliation[a]{Beijing Key Laboratory of Optical Detection Technology for Oil and Gas, China University of Petroleum-Beijing, Beijing 102249, China}
\affiliation[b]{Basic Research Center for Energy Interdisciplinary, College of Science, China University of Petroleum-Beijing, Beijing 102249, China}
\affiliation[c]{Institute for Theoretical and Mathematical Physics, Lomonosov Moscow State University, Moscow 119991, Russia}
\emailAdd{jiaqichen@cup.edu.cn}
\emailAdd{tyixiao@itmp.msu.ru}

\abstract{
Recent studies show that in addition to the flat spacetime, Integration-by-parts (IBP) relations and differential equations can also provide a systematic approach to
Feynman integrals in de Sitter spacetime. \texttt{dSIBP} is a \textsc{Mathematica} package
for the automated construction of IBP systems and differential equations. For dS integral families with massive and massless
propagators, it generates a complete set of IBP seeds, assembles linear systems for external reduction,
and constructs differential equations for the reduced master integrals. The implementation
combines equations of motion, remaining terms and parity constraints, and provides reliable checks.
We describe the conventions and public
interface through worked examples.
}

\begin{document}
\maketitle

\section{Introduction}
\label{sec:intro}

Correlation functions in de Sitter (dS) spacetime describe the quantum fluctuations relevant to
inflation. In the Schwinger--Keldysh formulation, their Feynman diagrams involve integrations over
vertex times and, at loop level, spatial loop momenta \cite{Chen:2017ryl}. Massive propagators
introduce Hankel functions into these integrals. Generalized integration-by-parts (IBP) relations
and differential equations (DE) offer a systematic way to organize this calculation
\cite{Chen:2023iix,Chen:2024glu,Chen:2026dqp,Arkani-Hamed:2023kig}.

In flat-spacetime QFT, IBP relations reduce integral families to master integrals
\cite{Chetyrkin:1981qh,Laporta:2000dsw}. Differentiating these master integrals and reducing the result gives
closed differential equations in the kinematic variables
\cite{Kotikov:1990kg,Kotikov:1991pm,Bern:1993kr,Gehrmann:1999as}.
A suitable basis can simplify the equations, in particular to canonical d $\log$ form
\cite{Henn:2013pwa}. This approach is supported by reduction programs such as FIRE
\cite{Smirnov:2008iw,Smirnov:2014hma,Smirnov:2019qkx,Smirnov:2025prc} and Kira
\cite{Maierhofer:2017gsa,Klappert:2020nbg}, rational-function reconstruction tools \cite{Magerya:2022hvj,Peraro:2019svx,Maier:2024djk,Belitsky:2023qho} (such as
Rational Tracer, FiniteFlow and related tools), and numerical DE solvers such as DiffExp
\cite{Hidding:2020ytt}. The auxiliary mass flow method and its implementation AMFlow provide
another example of turning reduction and differential equations into an automated numerical
calculation \cite{Liu:2017jxz,Liu:2022chg,Liu:2022mfb,Huang:2026rjb}.

For dS integrals, \cite{Chen:2023iix} generalizes IBP reduction to the cases whose integrands involve Hankel functions. The paper also gives universal formulas for recursive reduction and d $\log$
differential equations for arbitrary tree-level cosmological correlators and their loop integrands.
\cite{Chen:2024glu} applies a power-series method to solve the differential equations of tree-level dS correlators, establishes the factorization property of the solutions and analyzes their boundary conditions.
At loop level, \cite{Chen:2026dqp} combines time and momentum IBP relations and organizes
the system into subsystems labeled by the parities of the master integral indices, which comes from the propagators. Its massive bubble
example demonstrates the reduction to master integrals and the construction of d $\log$
differential equations. Together these results provide the basis for a systematic computational
approach to dS integrals (see also
\cite{Gasparotto:2024bku,Liu:2024str,Baumann:2026atn,De:2024zic,Grimm:2024tbg,Grafe:2026avi}).

\texttt{dSIBP} automates the generation of IBP relations in such systems, especially at loop level. Starting from vertices, propagators and momentum
assignments, it builds an integral family and generates a complete set of IBP seeds. The user chooses the ranges of integral indices and
passes the resulting linear system to an external IBP reducer such as Kira. After importing the reduction results, the package provides commands to differentiate with respect to the kinematic variables and construct differential equations for the chosen master integrals. Note that \texttt{dSIBP} does not launch an external reduction package or provide a numerical solver
for the resulting DE.

\texttt{dSIBP} also equips a \texttt{timeOnly} representation, which is available for diagrams without loop
momentum integrals and uses the same IBP and derivative metadata. Comparisons with analytic results and with MadStree
\cite{Chen:2026iwf} test the implementation as described in
section~\ref{sec:validation}.

This paper is organized as follows. Section~\ref{sec:notation} introduces the integral representation and the conventions needed for
Feynman diagrams. Section~\ref{sec:using} follows the calculation from initialization to
differential equations. Section~\ref{sec:examples} presents the examples and validation results. The public
functions are listed in Appendix~\ref{app:api}.

This package is available at \url{https://github.com/J7C/dSIBP-Packages}.

\section{Notation and conventions}
\label{sec:notation}

A Feynman diagram fixes the integral family on which the IBP and derivative operators act.
This section specifies the propagator states, index conventions and kinematic variables needed
to interpret that family and its subsectors.

\subsection{Schwinger--Keldysh Feynman rules}
\label{sec:skrules}
Correlation functions are evaluated in the Schwinger--Keldysh (SK) formalism, where the operator
product is ordered along a closed time path. A graph consists of vertices $v\in V$, internal lines
$e\in E$, and external legs. Each vertex is placed on either branch of the closed time path and is
assigned conformal time $\tau_v<0$, contour sign $\sigma_v=\pm1$, and signed total external energy
$E_v$. We first state the complete set of rules in this physical form. Coupling constants, external
wavefunction normalizations, and the overall momentum-conservation delta are suppressed; they
multiply the final correlation function and do not alter the loop-integration family.

\begin{enumerate}
\item \emph{Internal-line rule.} For a scalar of mass \(m\) propagating in \(d\) spatial
dimensions, the positive-frequency mode is
\begin{equation}
u_k(\tau)=-\frac{\mathrm{i}\sqrt{\pi}}{2}
e^{\mathrm{i}\pi(\nu/2+1/4)}\mathcal H^{(d-1)/2}
(-\tau)^{d/2}H^{(1)}_\nu(-k\tau),
\qquad
\nu^2=\frac{d^2}{4}-\frac{m^2}{\mathcal H^2},
\label{eq:bulkMode}
\end{equation}
where $k=|\bm k|$ and $\mathcal H$ is the Hubble parameter. The normalization follows the
convention of \cite{Chen:2026iwf}. Now define
\begin{equation}
G_>(k;\tau_1,\tau_2)=u_k(\tau_1)u_k^*(\tau_2),
\qquad
G_<(k;\tau_1,\tau_2)=u_k^*(\tau_1)u_k(\tau_2).
\label{eq:bulkCorrelators}
\end{equation}
For an oriented internal line \(e=(u,v)\) with contour signs \(\sigma_u,\sigma_v\), the bulk-to-bulk propagator \(G_{\sigma_u\sigma_v}\) is given by Feynman rules:
\begin{equation}
\begin{split}
G_{++}&=G_>\theta(\tau_1-\tau_2)+G_<\theta(\tau_2-\tau_1),\\
G_{--}&=G_<\theta(\tau_1-\tau_2)+G_>\theta(\tau_2-\tau_1),\\
G_{+-}&=G_<,\qquad G_{-+}=G_>.
\end{split}
\label{eq:skPropagators}
\end{equation}
A bulk-to-bulk propagator is called a \emph{time-ordered propagator} when its two contour signs agree and a \emph{Wightman propagator} when
they differ. Thus \(G_{++}\) and \(G_{--}\) are time-ordered propagators, while \(G_{+-}\) and
\(G_{-+}\) are Wightman propagators. Only the time-ordered propagators contain a time-ordering step
function; a Wightman propagator is
a pure product of two modes.

\item \emph{External-line rule.} Every external leg contributes a bulk-to-boundary propagator. For a massless external field, this factor is
\begin{equation}
G_\pm(\tau;k)\propto(1\mp \mathrm{i}k\tau)e^{\pm\mathrm{i}k\tau}.
\label{eq:bulkBoundary}
\end{equation}
Its sign is tied to the branch of the vertex at which the leg ends.

\item \emph{Vertex time integration.} Every vertex supplies one time integration over \(\tau_v\in(-\infty,0)\).

\item \emph{Loop-momentum integration.} Momentum conservation is imposed at every vertex on the spatial direction. An
internal line carries the momentum dictated by the external and loop momenta attached to that
vertex, and each independent loop momentum is integrated over its \(d\)-dimensional space.
\end{enumerate}

The rules above, together with momentum conservation, are the Feynman rules from which the
integral family is constructed.

\subsection{The integral family}
\label{sec:family}

An internal propagator is classified as massive
or massless in addition to being time-ordered or Wightman, giving the four types
$\langle$\texttt{massive, massless}$\rangle\times\langle$\texttt{time-ordered, Wightman}$\rangle$.
The assignments \(++\) and \(--\) belong to the same time-ordered class, and \(+-,-+\) to the same Wightman
class, but they still retain the branch dependence shown in \eqref{eq:skPropagators}.

Every propagator
contained in at least one loop carries loop momentum and contributes a denominator of the
integral family. The union of all such denominators is the \emph{loop skeleton} of the graph. A
propagator that does not lie on any loop is a \emph{tree propagator}, also called a \emph{bridge}; such a propagator has
no loop-momentum denominator.

\begin{figure}[t]
\centering
\includegraphics{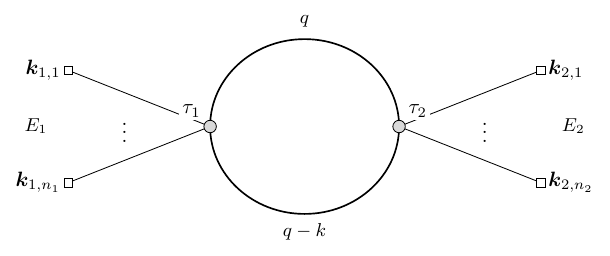}
\caption{Feynman rules of a de Sitter bubble with general contour signs, drawn as gray vertices.
The vertex energies \(E_1,E_2\) and the conformal times \(\tau_1,\tau_2\) are written at the
vertices. The external legs end in open squares and carry the momenta
\(k_{1,1},\dots,k_{1,n_1}\) and \(k_{2,1},\dots,k_{2,n_2}\). The two internal propagators carry
the loop momenta \(q\) and \(q-k\).}
\label{fig:skrules}
\end{figure}

The IBP reduction on theta functions pinches the corresponding $++$ or $--$ propagators and
produces new master integrals, called remaining terms \cite{Chen:2023iix}. A sector $s$ specifies
which propagators remain and which have been pinched, and the \emph{top sector} is the diagram before any pinch with its propagator order
fixed for the whole family. A sector $s_1$ coming from pinching propagators in the sector $s_0$ is called a \emph{subsector} of $s_0$, while $s_0$ is called a \emph{super sector} of $s_1$. These remaining terms are divided into
different subsectors according to the number of pinched propagators and their position in the
Feynman diagram.

Based on the clarifications above, the whole integral family has the following expression 
\begin{equation}
\begin{split}
J_s={}&N_s\!\int\prod_{\ell=1}^{L}d^dq_\ell
\int_{-\infty}^{0}\prod_{v\in V_s}d\tau_v\;
\prod_{v\in V_s}(-\tau_v)^{A_v}e^{-\mathrm{i}\sigma_v E_v\tau_v}\\
&\times\prod_{e\in\mathcal C}\xi_e^{-B_e}
\prod_{e\in U_s}\mathcal B_e^{(n_{e,1},n_{e,2})}
(\xi_e;\tau_u,\tau_v)\;
\prod_r\rho_r^{\,i_r}\,.
\end{split}
\label{eq:family}
\end{equation}
The objects entering this expression are defined as follows.
\begin{enumerate}
\item \emph{Integration data.} The graph has $L$ independent loops, and
\(q_1,\dots,q_L\) are the corresponding spatial loop momenta integrated over the
$d$-dimensional space. The set \(V_s\) contains all vertices of the sector. Each vertex
carries one time integral over \((-\infty,0)\). 

\item \emph{Vertex time factors.} The exponent
\(A_v=a_v+a_{0v}\) is the effective vertex-time power. The integer \(a_v\) is the IBP shift,
whereas \(a_{0v}\) is the declared zero point. The phase
\(e^{-\mathrm{i}\sigma_vE_v\tau_v}\) comes from the external legs and the massless Wightman propagators.

\item \emph{Loop denominators.} The set \(\mathcal C\) contains all propagators lying on a
loop of the root Feynman diagram and does not depend on the sector. The quantity \(\xi_e\) is
the magnitude of the loop momentum carried by \(e\). We define the sector-dependent power
\begin{equation*}
B_e=
\begin{cases}
b_e+b_{0e}, & e\ \text{unpinched},\\
b_{S,e}+b_{S0,e}, & e\ \text{pinched},
\end{cases}
\end{equation*}
where \(b_{S0,e}\) is the corresponding sector zero point. Thus the factor
\(\xi_e^{-B_e}\) supplies either the original loop denominator or its pinched residual power.
For a two-line bubble, for example, the unpinched loop denominators form
\(\xi_1^{-(b_1+b_{01})}\xi_2^{-(b_2+b_{02})}\).

\item \emph{Schwinger--Keldysh blocks.} The set \(U_s\) contains all unpinched propagators in
sector \(s\). A pinched propagator has no block in \(U_s\); its contribution is carried by the
corresponding power in \(B_e\) and by \(N_s\). For \(e=(u,v)\),
we denote the normalized SK block used by the IBP system by
\(\mathcal B_e^{(n_{e,1},n_{e,2})}\). Let \(x_u=-\xi_e\tau_u\), \(x_v=-\xi_e\tau_v\),
\(\theta_{uv}=\theta(\tau_u-\tau_v)\), and define
\begin{equation}
\begin{aligned}
g_>^{(n_1,n_2)}&=
h^{(1)}(\nu_e,n_1;x_u)h^{(2)}(\nu_e,n_2;x_v),\\
g_<^{(n_1,n_2)}&=
h^{(2)}(\nu_e,n_1;x_u)h^{(1)}(\nu_e,n_2;x_v),
\end{aligned}
\label{eq:reducedCorrelators}
\end{equation}
with the two-state functions \(h^{(\alpha)}\) defined in \eqref{eq:hDef}. The normalized
propagator is
\begin{equation}
\begin{aligned}
\mathcal G_{++}^{(n_1,n_2)}
&=g_>^{(n_1,n_2)}\theta_{uv}+g_<^{(n_1,n_2)}\theta_{vu},\\
\mathcal G_{--}^{(n_1,n_2)}
&=g_<^{(n_1,n_2)}\theta_{uv}+g_>^{(n_1,n_2)}\theta_{vu},\\
\mathcal G_{+-}^{(n_1,n_2)}
&=g_<^{(n_1,n_2)},\qquad
\mathcal G_{-+}^{(n_1,n_2)}=g_>^{(n_1,n_2)}.
\end{aligned}
\label{eq:normalizedBlocks}
\end{equation}
For a massive propagator,
\(\mathcal B_e^{(n_1,n_2)}=\mathcal G_{\sigma_u\sigma_v}^{(n_1,n_2)}\). For a massless time-ordered
propagator it is the block \(F_{n_1n_2}\) defined in \eqref{eq:masslessQuotient}, while for a massless
Wightman propagator
\(\mathcal B_e^{+-}=e^{\mathrm{i}\xi_e(\tau_u-\tau_v)}\) and
\(\mathcal B_e^{-+}=e^{-\mathrm{i}\xi_e(\tau_u-\tau_v)}\).

Note that \eqref{eq:normalizedBlocks} does not include the common mode-function normalization
\begin{equation}
-\frac{\mathrm{i}\sqrt{\pi}}{2}
e^{\mathrm{i}\pi(\nu/2+1/4)}\mathcal H^{(d-1)/2},
\nonumber
\end{equation}
the powers of \(k\) and \(\tau\), or other source-level overall constants in
\eqref{eq:bulkMode}. These factors are carried by the line-dependent normalization and, after
pinches, by \(N_s\). Thus \(\mathcal B_e\), rather than the full physical propagator
\(G_{\sigma_u\sigma_v}\), is the object entering the IBP integral family. The endpoint-state labels
are \(n_1,n_2\in\{0,1\}\), where \(n=0\) selects the undifferentiated \(h\) function and \(n=1\)
selects its first \(x\)-derivative, as in \eqref{eq:hDef}.

\item \emph{Irreducible scalar products.} The \(\rho_r\) are the irreducible scalar products
(ISPs), and \(i_r\in\mathbb Z\) are their integer powers. Positive powers are numerator powers,
while negative powers give additional denominators. Their construction is discussed in
section~\ref{sec:kinematics}. For example, if \(\rho_1=\operatorname{sp}[q_1,k]\), then
\(\rho_1^{i_1}\) contributes a numerator for \(i_1>0\) and an additional denominator for
\(i_1<0\).

\item \emph{Sector normalization.} The factor \(N_s\) contains constants, normalizations
generated by pinches, and the powers of non-loop momentum magnitudes. Pinching tree propagators of the super sector of $s$
contributes momentum magnitude powers through \(N_s\). Kinematic derivatives act on \(N_s\) as well
as on the remaining integrands. Its construction and the conversion of remaining terms are described later.

\end{enumerate}

As an explicit example, the two-line bubble of figure~\ref{fig:skrules} has
\begin{equation}
\begin{split}
J_{\mathrm{bubble}}={}&N_s\int\mathrm d^dq\int_{-\infty}^{0}
\mathrm d\tau_1\mathrm d\tau_2\,(-\tau_1)^{A_1}(-\tau_2)^{A_2}\\
&\times e^{-\mathrm i\sigma_1E_1\tau_1-\mathrm i\sigma_2E_2\tau_2}
\xi_1^{-B_1}\xi_2^{-B_2}\\
&\times\mathcal B_1^{(n_{1,1},n_{1,2})}
\mathcal B_2^{(n_{2,1},n_{2,2})}\rho_1^{i_1}.
\end{split}
\label{eq:bubbleExample}
\end{equation}

In the package this integral family is written in the three-slot schema for loop diagrams
\begin{verbatim}
J[aList, linePacks, ispList]
\end{verbatim}
Here \texttt{aList} is the ordered list of vertex shifts \(a_v\),
\texttt{ispList} contains the integer powers \(i_r\), and \texttt{linePacks} contains one pack per
declared propagator in the order fixed by the declaration key \texttt{"lines"}. The possible pack
shapes are summarized below.
\begin{center}
\small
\begin{tabular}{@{}p{5.0cm}lp{0.40\textwidth}@{}}
\toprule
Propagator & Pack & Interpretation \\
\midrule
Loop massive time-ordered/Wightman & $\{b_e,n_{e,1},n_{e,2}\}$ & power and two endpoint states \\
Loop massless time-ordered & $\{b_e,n_{e,1},n_{e,2}\}$ & two representatives after the quotient \\
Loop massless Wightman & $\{b_e,0,0\}$ & one exponential block, no state bit \\
Tree propagators& $\{\texttt{"F"},n_{e,1},n_{e,2}\}$ & no denominator; magnitude in $N_s$ \\
Pinched loop propagator & $\{b_{S,e}\}$ & residual loop-momentum power \\
Pinched tree propagator & $\{\texttt{"F"}\}$ & placeholder slot for the residual tree momentum; power in $N_s$ \\
\bottomrule
\end{tabular}
\end{center}

The first slot is \(b_e\) for an indexed loop propagator and the string \texttt{"F"} for a tree
propagator; \texttt{"F"} is a placeholder slot for the residual tree momentum and carries no power
index, the actual momentum-magnitude power staying in \(N_s\). Thus a time-ordered pack
\(\{b_e,n_{e,1},n_{e,2}\}\) contributes
\(\xi_e^{-(b_e+b_{0e})}\mathcal B_e^{(n_{e,1},n_{e,2})}\), a pinched pack \(\{b_{S,e}\}\)
contributes \(\xi_e^{-(b_{S,e}+b_{S0,e})}\), and a tree pack
\(\{\texttt{"F"},n_{e,1},n_{e,2}\}\) contributes
\(\mathcal B_e^{(n_{e,1},n_{e,2})}\) with its momentum-magnitude power retained in \(N_s\).
For a massless time-ordered propagator only the two representatives \(F_{00}\) and \(F_{10}\) are
independent; for a massless Wightman propagator the two state slots are fixed to zero.

\begin{figure}[t]
\centering
\includegraphics{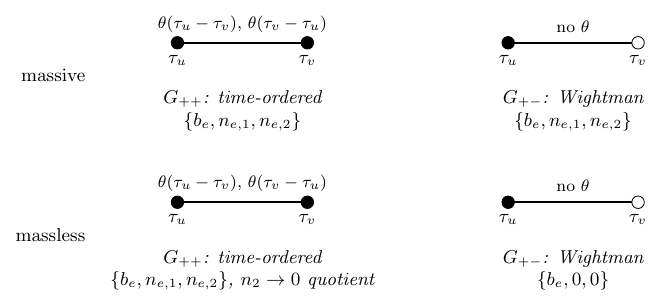}
\caption{The four types of internal propagator and their packs. The fill color of a vertex
encodes its contour sign: black is \(+\) and white is \(-\). The conjugate assignments \(--\) and
\(-+\) give the same two types. Massive time-ordered, massive Wightman and massless time-ordered propagators carry
\(\{b_e,n_{e,1},n_{e,2}\}\), with the massive vertex states restricted to \(\{0,1\}\). A massless
Wightman propagator is a single exponential product, so it carries \(\{b_e,0,0\}\). A massless time-ordered
propagator has two independent states with canonical second state power zero.}
\label{fig:linetypes}
\end{figure}

\paragraph{Massive propagators.}
\label{sec:dynamics}
Every massive propagator is assigned a two-dimensional mode-function space described by the
function system
\begin{equation}
f''+P(x)f'+Q(x)f=0,
\qquad x=-\xi\tau.
\label{eq:functionSystem}
\end{equation}
The default \texttt{"h"} preset uses the two-state functions of Eq.~(2.8) in
\cite{Chen:2026iwf},
\begin{equation}
h^{(\alpha_e)}(\nu_e,n_e;x)=
\begin{cases}
x^{\nu_e}H^{(\alpha_e)}_{\nu_e}(x), & n_e=0,\\[2pt]
\partial_x h^{(\alpha_e)}(\nu_e,0;x), & n_e=1,
\end{cases}
\qquad \alpha_e=1,2,
\label{eq:hDef}
\end{equation}
with \(x=-\xi_e\tau\), which is a deformed Hankel function with better properties. It has
\begin{equation}
P_{\mathrm h}(x)=\frac{1-2\nu_e}{x}, \quad Q_{\mathrm h}(x)=1.
\end{equation}
The built-in \texttt{"H"} preset is for the original Hankel function, which has
\begin{equation}
P_{\mathrm H}(x)=\frac{1}{x}, \quad Q_{\mathrm H}(x)=1-\frac{\nu_e^2}{x^2}.
\end{equation}
A massive propagator $e$ has two endpoint-state labels $n_{e,1},n_{e,2}\in\{0,1\}$, which select the mode function and its first derivative. In the default \texttt{"h"} system, the second-order equation of motion at $n_{e,i}=2$ gives
\begin{equation}
\begin{split}
J\big[\cdots,n_{e,i}=2,\cdots\big]
={}&-J\big[\cdots,n_{e,i}=0,\cdots\big]\\
&-(1-2\nu_e)\,J\big[\cdots,n_{e,i}=1,\,a_{e,i}\!\to\!a_{e,i}\!-\!1,\,b_e\!\to\!b_e\!+\!1,\cdots\big],
\end{split}
\label{eq:eom}
\end{equation}
so every massive endpoint state reduces to the representative set \(\{0,1\}\). For a tree
propagator the second term instead carries an explicit factor \(\xi_e^{-1}\) and only shifts
\(a_{e,i}\to a_{e,i}-1\), since there is no propagator power to shift. The package applies this
reduction by construction, so the generated seeds contain only \(n_{e,i}\in\{0,1\}\).
Every baseline formula involving \(\nu_e\), including \eqref{eq:eom} and the pinched-term shifts below,
is written with \(\nu_e\) equal to the Hankel order of \eqref{eq:bulkMode} as supplied in the
input. The initialization option \texttt{NuConvention} of section~\ref{sec:init} selects this
writing; the alternative \texttt{"Negative"} flips the sign of \(\nu_e\) in every baseline formula
and reproduces the equations of the opposite writing. Appendix~\ref{app:nu} lists the affected
shapes.

The optional line key \texttt{"functionSystem"} allows the function system to be replaced by a
user-defined association
\begin{verbatim}
<|"variable" -> x, "P" -> ..., "Q" -> ...,
  "T" -> ..., "W" -> ...|>
\end{verbatim}
Here \texttt{"variable"} defaults to \texttt{x}, \texttt{"T"} defaults to the identity matrix, and
\texttt{"P"}, \texttt{"Q"} and \texttt{"W"} define the equation of motion and its Wronskian. The
coefficient functions must admit the finite Laurent-shift representation used by the generators.
Let \(T\) be an invertible \(2\times2\) basis transformation acting on
\((f,f')^{\mathsf T}\). For two independent solutions \(f_1,f_2\), the Wronskian in the original
derivative basis is
\begin{equation}
W[f_1,f_2]=f_1f_2'-f_1'f_2.
\label{eq:wronskian}
\end{equation}
The built-in massive presets use
\begin{equation}
W_{\mathrm h}(x)=
-\frac{4\mathrm i}{\pi}e^{\pi\operatorname{Im}\nu_e}x^{2\bar\nu_e-1},
\qquad
W_{\mathrm H}(x)=
-\frac{4\mathrm i}{\pi}e^{\pi\operatorname{Im}\nu_e}\frac{1}{x},
\label{eq:presetWronskians}
\end{equation}
where
\begin{equation*}
\bar\nu_e=
\begin{cases}
\nu_e, & \texttt{NuConvention}\to\texttt{"Positive"},\\
-\nu_e, & \texttt{NuConvention}\to\texttt{"Negative"}.
\end{cases}
\end{equation*}
At
initialization the package compiles \(A_T=T'T^{-1}+TA_0T^{-1}\), where
\(A_0=\begin{pmatrix}0&1\\-Q&-P\end{pmatrix}\), and \(W_T=\det(T)W\). It checks
\(W'+PW=0\) and rejects singular or inconsistent input. The compiled derivative terms are the
equation-of-motion shifts acting on the state labels. The compiled \(W_T\) supplies the Wronskian
terms used for pinches. When an IBP reduction differentiates the theta functions and produces
remaining terms, it is exactly this compiled \(W_T\), and hence
\eqref{eq:presetWronskians} for the built-in massive presets, that fixes the pinch
coefficient. Ordinary derivatives of
kinematic coordinates never invoke \(W_T\).
If an explicit \texttt{"WT"} is supplied, it is only checked against \(\det(T)W\). The IBP
generators and derivative operators all read the same compiled function system.
Example~02 supplies a bare Hankel system through this interface.

\paragraph{Massless propagators.}
A massless time-ordered propagator $e$ between the ordered vertices $u$ and $v$ has two independent
states. Let $\Delta=\tau_u-\tau_v$ and let $q$ be the magnitude of its momentum. Its kernel is
built from the two functions
\begin{equation}
\begin{split}
M_0={}&\theta(\Delta)\,e^{-\mathrm{i}\sigma q\Delta}
      +\theta(-\Delta)\,e^{+\mathrm{i}\sigma q\Delta},\\
M_1={}&-\theta(\Delta)\,e^{-\mathrm{i}\sigma q\Delta}
      +\theta(-\Delta)\,e^{+\mathrm{i}\sigma q\Delta},
\end{split}
\label{eq:masslessBlocks}
\end{equation}
where $\sigma$ is the common contour sign of the two endpoints, so $\sigma=+1$ on $++$ and
$\sigma=-1$ on $--$. Exchanging the two vertices leaves $M_0$ unchanged and flips the sign of $M_1$.
Write $F_{n_1n_2}$ for its propagator block with
$(n_{e,1},n_{e,2})=(n_1,n_2)$. The package works with the quotient
\begin{equation}
F_{00}=M_0,\qquad F_{10}=M_1,\qquad
F_{01}=-F_{10},\qquad F_{11}=-F_{00},
\label{eq:masslessQuotient}
\end{equation}
so only the two representatives $F_{00}$ and $F_{10}$ are independent. The pack of such a
propagator nevertheless keeps both endpoint-state labels, each taking the values \(0\) and \(1\).
Keeping both labels lets the symmetry relations act directly on the ordered endpoint positions,
without changing the line-pack representation at intermediate steps. For a line with ordered
endpoints \(e=(u,v)\), \(n_{e,1}\) is defined relative to \(u\), so \(F_{10}\) is the positive
antisymmetric representative and \(F_{01}=-F_{10}\). Exchanging the two endpoints therefore
flips \(n_{e,1}\). Seeds are generated for the two representatives only, and a state produced by
a derivative is reduced with \(F_{n_1 1}=-F_{1-n_1 0}\), equivalently \(n_{e,2}\to0\). This is an
exact algebraic identification, not a numerical sampling rule. A massless Wightman propagator is
instead a single pure-exponential block and therefore carries the state labels \((0,0)\).
Several time-ordered propagators may join the same pair of vertices; we call the set of time-ordered propagators
sharing two endpoints a banana sub-diagram. A propagator belonging to a banana sub-diagram is one
of the multiple propagators that share the two endpoints and is called a banana propagator
hereafter. Banana massless time-ordered propagators obey one further algebraic relation: their kernel
product depends on the endpoint states only through the total parity $\sum_e n_{e,1}\bmod 2$,
because the mixed time-ordering terms of two banana members vanish identically. The package
folds such a banana sub-diagram onto a canonical representative in which the first line carries
the parity and every other member is in the state $(0,0)$.
Appendix~\ref{app:fold} derives this fold in the original index representation and fixes the sign
picked up from endpoint exchanges.

\paragraph{Massless tadpoles.}
A massless propagator whose two endpoints coincide, a massless tadpole, enters a family in two
ways: declared with equal endpoints in the input, or produced when a pinch merges the
endpoints while a massless member of the banana sub-diagram survives. Both are converted to the
pinched propagator representation: declared tadpoles at initialization, and pinch
survivors when the sector topology is built. The reason is the coincident kernel: with $\Delta=0$,
$M_0=2\theta(0)=1$ and $M_1=0$, so the undifferentiated block is the constant $1$ and the
antisymmetric states vanish identically. Only the residual momentum power survives, so the pack
becomes the single slot $\{b_{S,e}\}$ with $b_{S,e}=b_e$ and an unshifted zero point, as for any
massless pinch, and no endpoint-state labels are kept. In the sector inventory, surviving
massless members of a pinched banana are folded into the pinched set, and distinct
pinch routes that reach the same pinched set define a single sector. A massive tadpole
is not subject to this conversion: it keeps its three-slot pack and the loop-reversal relation of
section~\ref{sec:seeds}.

\paragraph{Remaining terms: pinched propagator and sector normalization.}
Differentiating the time-ordering factors produces remaining terms \cite{Chen:2023iix}. For the default massive
and massless time-ordered blocks, a nonzero elementary pinch requires $n_{e,1}+n_{e,2}=1$.
The package merges the endpoints, removes the corresponding block, and maps the time powers,
momentum powers and zero points to the subsector. For a loop propagator in the default
\texttt{"h"} system, for example, the merged time power is $a_u+a_v-1$, its zero point is
$a_{0u}+a_{0v}+2\nu_e$, and the residual sector momentum powers are $b_{S,e}=b_e+1$ and
\(b_{S0,e}=b_{0e}-2\nu_e\). Under the \texttt{"Negative"} convention of appendix~\ref{app:nu} the
two $\nu_e$ shifts reverse their signs. For a massless pinch the time powers add without
these shifts, and
$b_{S,e}=b_e$. The time-ordered propagators of a banana sub-diagram share one
time-ordering boundary and must be differentiated as a product. For each banana sub-diagram
\(B\), the expansion into pinched terms contains nonempty odd subsets \(Q\subseteq B\) with combinatorial
factor
\(2^{1-|Q|}\). Each term merges the \emph{pinched vertex pair}, the two vertices shared by the
banana sub-diagram, once. If
\(c^{\mathrm{raw}}_{sQ}\) is its coefficient in the unnormalized integral convention, including
the combinatorial factor and elementary coefficient of the remaining term, its contribution in the
formula with normalization factors is
\begin{equation}
\sum_{\substack{\emptyset\neq Q\subseteq B\\ |Q|\ \mathrm{odd}}}
c^{\mathrm{raw}}_{sQ}\,\frac{N_s}{N_{t_Q}}\,J_{t_Q}\,.
\label{eq:pinched}
\end{equation}
Here \(t_Q\) denotes the subsector with all index and zero-point shifts applied. The ratio of
normalizations is essential: the sub-normalization already contains the pinch constants
and non-loop momentum magnitude powers. The package performs this conversion when assembling IBP
relations, so these factors must not be inserted into the final integrals a second time.
Appendix~\ref{app:banana} derives the merger of the shared time-ordering vertices pair and the
odd-subset structure behind \eqref{eq:pinched} for an arbitrary banana sub-diagram.

\subsection{Kinematics}
\label{sec:kinematics}
Let $N_I$ count the internal propagators, $N_V$ count the vertices and $C$
the connected components. Then the loop number $L$ can be written as
\begin{equation}
L=N_I-N_V+C\,.
\label{eq:loopnumber}
\end{equation}
Together with $K$ independent
loop-related external momenta, the loop number fixes the number of scalar products,
\begin{equation}
N_{\mathrm{sp}}=\frac{L(L+1)}{2}+LK\,.
\label{eq:nsp}
\end{equation}
The user declares two separate lists. The first list,
\texttt{loopExternalMomenta} $\{k_1,\dots,k_K\}$, defines the independent scalar-product
parameters. By default,
\begin{equation*}
\texttt{ss}ij=\sqrt{\mathrm{sp}[k_i,k_j]}.
\end{equation*}
The first \texttt{s} of \texttt{ss}ij denotes the square root and the second denotes the scalar
product, so \texttt{ss}ij is the square root of the $(i,j)$ entry of the loop Gram matrix.
The second list, \texttt{independentExternalMomenta}, defines the non-loop momentum magnitudes
$|k_{Ei}|$ that appear in vertex energies and in the masses of tree propagators. These
magnitudes are
parameterized by $\texttt{sE}i$. Only
magnitudes are coordinates. The package never outputs $\mathrm{sp}[k_{Ei},k_{Ej}]$, and it never
eliminates intermediate magnitudes by a cosine rule. Magnitudes that depend on the declared ones
are stored as explicit bindings. They still contribute to vertex phases through the chain rule.

User-defined ISPs are declared as \texttt{name/expr/range} associations, for example
\begin{verbatim}
<|"name" -> rho1, "expr" -> sp[q1, k], "range" -> {0, 1}|>,
\end{verbatim}
and they must form an invertible coordinate system on the space of scalar products together with
the propagator squares. The definition zero point of an ISP is fixed at $0$: positive powers are
numerator powers, and a negative power means an extra denominator.

The kinematic coordinates of a case are fixed by rules that express the basic scalar products in
terms of chosen symbols. A rule has the form
\texttt{sp[k1,k1] -> u\textasciicircum2}, and a rule can bind a whole energy as well, as
\texttt{sp[k1,k1] -> (E0-sE2)\textasciicircum2} does in the mixed example of
section~\ref{sec:examples}. The set of these rules is the kinematic system of the case, and its
symbols are exactly the variables that later derivatives and scaling checks use. The two-stage
interface \texttt{DSKinematics}/\texttt{DSInit}
validates this system and reports an exact, over-complete or under-complete
state. Only the exact state opens derivative generation.

\subsection{Parity structure}
\label{sec:parity}

For the massive \texttt{"h"} and \texttt{"H"} presets and the massless systems, the discrete
invariant of the dS IBP system is the parity of $b_e+n_{e,1}+n_{e,2}$ on a propagator.
The package does not automatically extend this parity support to general custom function systems.
For every propagator $e$ that carries a propagator power
$b_e$, that is for every propagator of the
loop skeleton, the quantity
\begin{equation}
p_e \;=\; b_e+n_{e,1}+n_{e,2} \pmod 2
\label{eq:parity}
\end{equation}
is conserved by every time generator and every loop-momentum generator. A massless Wightman propagator has
$n_{e,1}=n_{e,2}=0$, so its parity is just $b_e$. A tree propagator carries no propagator power, so
no parity is defined for it. The parity vector
$(p_e)_e$ of the top sector labels the closed subsystems of the IBP system of a family, and for a
diagram whose propagators all lie on the loop skeleton this is exactly the split stated by \cite{Chen:2026dqp}.

The default \texttt{"h"}
and \texttt{"H"} pinches of a massive or massless propagator leave the parity
unchanged. The package therefore does not apply a constraint to a sector
below the top sector in its original form. It rewrites the constraint in the powers of that sector,
using the power and zero-point shifts of its pinched propagators. When the rewrite is not integral, the
parity filter is disabled.

For families with
user-provided \texttt{parityConstraints} the package keeps only the seeds whose transported constraint
vanishes. A constraint is written in the powers of \eqref{eq:parity}, for example
$b_1+n_{1,1}+n_{1,2}\equiv0$ for the propagator with declaration index $1$; the declaration spells
the same powers as \texttt{b[1]}, \texttt{n[1,1]} and \texttt{n[1,2]}. ISP indices receive no
parity filter.

\subsection{Tree-level and time-only families}
\label{sec:timeonly}

Tree-level quantities require no loop momenta, loop skeleton or ISPs. The complete tree-level
time-integral family, sector labels and mode-function state conventions are given in \cite{Chen:2026iwf}; we use those conventions when referring to the tree-level
track and do not repeat its reduction rules here.

For diagrams without loops the package exposes a \texttt{timeOnly} family with the unique public
representation
\begin{verbatim}
J[sectorKey_String, timeShifts_List, stateBits_List]
\end{verbatim}
The \texttt{sectorKey} records, for each propagator of the root
topology, whether it is unpinched (1)
or pinched (0). The list \texttt{timeShifts} gives the time powers $a_v$ of the vertices. The
list \texttt{stateBits} gives one bit for each massive vertex state or massless quotient state, in
the fixed order of the root propagators. There are no momentum generators, no propagator powers
and no ISPs. The magnitude powers of tree propagators remain in $N_s$
and still enter derivative generation. This mode is the default for tree diagrams. It can also
be selected explicitly for the time-integral part of a diagram with loops, with its internal
momentum magnitudes treated as parameters rather than integration variables. In the full loop
mode, a non-loop propagator remains a tree propagator in the three-slot family of
section~\ref{sec:family}. The general \texttt{DSSeeds}, \texttt{DSGenerateIBP} and
\texttt{DSLinear} route remains available for these families. The package also provides naive
time-IBP reduction and naive tree differential equations for certified massive-only tree
families, while MadStree \cite{Chen:2026iwf} provides the direct formula-based reduction and
d \(\log\) construction.

\section{Using the package}
\label{sec:using}

The workflow has three stages: construct the family and its IBP system, obtain an external
reduction, and use it to build differential equations. The commands below illustrate these
stages; complete runnable declarations and scripts are supplied with the examples in
section~\ref{sec:examples}. Load the distributed \texttt{dSIBP.m} with
\texttt{Get[packageFile]}, where \texttt{packageFile} is its full path. The
\texttt{main.wl} script of example~04 supplies the massive-bubble declaration used below, under the
name \texttt{caseInput}; \texttt{case} below stands for such an input association, and
\texttt{custom} for its custom kinematic rules, declared as \texttt{kinematicRules} in example~06.
Options follow the Wolfram convention, and \texttt{Options[function]} lists them at run time.
Two options are shared by several commands and are not repeated below: \texttt{ProgressReporting}
(\texttt{Automatic} by default) switches the console progress of the long-running commands on or
off, and \texttt{AuditLevel} (\texttt{"standard"} by default; \texttt{"full"} re-computes the
integrity, coverage, representation and residual audits) sets the audit depth of \texttt{DSGenerateIBP},
\texttt{DSLinear} and the
tree commands \texttt{DSTreeNaiveIBP} and \texttt{DSTreeNaiveDE}. The
remaining options and specification keys are introduced together with the commands that accept
them.

\subsection{Initialization}
\label{sec:init}

A case is a named diagram declaration. Initialization runs a
topology audit of the loop number,
the registry of vertices and propagators and the external-momentum routing. It then runs the
kinematic audit of
section~\ref{sec:kinematics}. It also checks the time-integration domain: if an integer shift of
the continuous indices drives the leading small-time powers of a vertex to a divergent boundary, the
run stops and names the vertices that need a symbolic regulator. The result is a context that every
later command consumes:
\begin{verbatim}
context = DSInit[case, KinematicRules -> custom];
\end{verbatim}
The optional two-stage interface \texttt{DSKinematics[case]} first returns a proposal with the
default rules and a selection template. The user can then inspect the kinematic coordinates
$\texttt{ss}ij,\texttt{sE}i$, the momentum-routing audit and the capability report before committing,
and \texttt{DSKinematics[case, rules]} audits a chosen set of rules in the same form.
Under-complete declarations abort the initialization. Over-complete ones allow symbolic seed
generation and diagnostic symbolic linear data but disable derivatives, the inner-form rewriting
and export to Kira. In the over-complete case common loop shifts are quotiented out through an
affine basis \texttt{effectiveLoopExternalMomenta}.

Initialization accepts the dedicated options
\begin{itemize}
\item \texttt{KinematicRules}: the custom rules above;
\item \texttt{NuConvention} (\texttt{"Positive"} by default): the sign convention in which the
baseline formulas involving $\nu_e$ are written, see section~\ref{sec:dynamics} and
appendix~\ref{app:nu}. \texttt{"Positive"} takes $\nu_e$ to be the Hankel order of
\eqref{eq:bulkMode}; \texttt{"Negative"} flips the sign of $\nu_e$ in every baseline formula. The
choice is resolved per massive line and frozen in the context; no later command accepts a
convention option;
\item \texttt{GenerateDerivativeMetadata} (\texttt{False} by default): pre-compute the metadata of
the family's derivative operators, which the written initialization files then include;
\item \texttt{WriteInitializationFiles} and \texttt{InitializationDirectory} (\texttt{False} and
\texttt{Automatic} by default): write the parsed topology, sectors and conventions as reloadable
files;
\item \texttt{OverwriteInitialization} (\texttt{False} by default): a directory holding a different
input is refused unless this option is set;
\item \texttt{RegisterAsCurrent} (\texttt{True} by default): \texttt{False} initializes without
replacing the current context, as \texttt{DSRedefineParameters} does internally.
\end{itemize}
The command \texttt{DSRedefineParameters} itself
re-initializes the same input with a complete new rule set. It accepts
\texttt{ProgressReporting} but not \texttt{AuditLevel}.

\subsection{Seeds and the IBP index ranges}
\label{sec:seeds}

The sectors of a family are ordered by their pinches. The \emph{top sector} retains
all propagator blocks; subsectors arise by pinching propagators and identifying their endpoints.
The indices of the top sector are the root indices used below.
\texttt{DSSeeds[context]} returns the complete set of IBP seed templates of the family, and
\texttt{DSAllSeeds} collects them into a list:
\begin{verbatim}
seedData = DSSeeds[context];
allSeeds = DSAllSeeds[seedData];
\end{verbatim}
Every sector
receives all the time generators of its vertices and all $L(L+K)$ loop-momentum generators.
For each sector, \texttt{DSSeeds} first enumerates the discrete state labels and applies the
equation of motion \eqref{eq:eom} to each general template. Massless time-ordered propagators are reduced
to the two quotient representatives of \eqref{eq:masslessQuotient}, the massless banana
sub-diagrams among them are
folded to the canonical parity form of appendix~\ref{app:fold}, and all remaining state labels
are either fixed to \(0,1\) or eliminated. The analytic seed templates still keep the general
continuous indices \(a_v,b_e,b_{S,e}\) and general ISP indices; these continuous indices are
expanded only afterward by \texttt{DSGenerateIBP}. Thus no state label is sampled as a continuous
index.

\paragraph{Symmetry relations.}
The package does not search for general graph symmetries. A symmetry must be declared by the user
in the optional \texttt{"symmetryRules"} key of the input. Each entry is a rule or delayed rule
acting on the integral, and may also map a representative to zero. For every equivalence class the
user must choose one canonical representative and supply only directed rules from the other
representatives to it; two opposite rules between the same representatives must not both be
listed. A typical rule therefore has the form
\begin{verbatim}
HoldPattern[(int_J /; exampleTopQ[int])] :> exampleTopCanonical[int]
\end{verbatim}
The massive-bubble example uses such rules for the vertex exchange and the exchange of its two
massive banana propagators. It assigns the list through
\texttt{"symmetryRules" -> exampleSymmetryRules0}; the helper predicates choose the canonical
representative before the replacement is applied.

After the equation of motion, the massless quotient and the banana fold, \texttt{DSSeeds} applies
the user rules
together with the built-in tadpole rules to each discrete template. A built-in tadpole rule is
available only for a self-loop whose reversal is a single loop-momentum inversion. If a user rule
can depend on integer indices, \texttt{DSGenerateIBP} applies the same canonicalization again after
the continuous-index values have been substituted. The atomic derivative through \texttt{ds} also
canonicalizes its output. The raw user rules are obtained by applying
\texttt{repSymmetry0} to \texttt{context["topology"]}. The function \texttt{symmetry} applies the
combined user and built-in rules once to an explicit expression.

The continuous indices are swept by \texttt{DSGenerateIBP}. The unified form
\begin{verbatim}
ibpData = DSGenerateIBP[allSeeds, {-1, 1}];
\end{verbatim}
requires that every root continuous index lies in the closed integer interval $[-1,1]$ \emph{after}
all shifts of the generated relations. The package inverts the per-index shift sets and derives the
narrower seed ranges automatically. The per-index form
\begin{verbatim}
ibpData = DSGenerateIBP[allSeeds,
   {a[v1], -1, 1}, {a[v2], 0, 1},
   {b[e1], 0, 2}, {b[e2], -1, 1}];
\end{verbatim}
must cover all root continuous indices exactly. ISP lower bounds are combined with the user target
by taking the larger lower bound, so the package never extends a range on its own in order to cover
numerator terms. Only complete systems may proceed to a formal reduction. \texttt{DSSeeds} carries
only \texttt{ProgressReporting}, not \texttt{AuditLevel}; \texttt{DSGenerateIBP} accepts the two
filters and the evaluation route described next.

\paragraph{Seed-envelope cutting.}
The option \texttt{SeedEnvelopeCut} defaults to \texttt{Automatic}. It filters the inferred seed
points before parity filtering and template substitution:
\begin{verbatim}
ibp0 = DSGenerateIBP[allSeeds, {-1, 1}, SeedEnvelopeCut->None];
ibp1 = DSGenerateIBP[allSeeds, {-1, 1}, SeedEnvelopeCut->Automatic];
ibp2 = DSGenerateIBP[allSeeds, {-1, 1}, SeedEnvelopeCut->{10, 4, 5, 4}];
\end{verbatim}
At a candidate point, let $C_+$ be the sum of its positive continuous slots and $C_-$ the
absolute sum of its negative slots, and let
\begin{equation}
  T = C_+ + C_-.
  \label{eq:seedEnvelopeCut}
\end{equation}
For $C_+\geq C_-$ the point is accepted when $T\leq Ex_1$ and $C_-\leq Ex_2$; for $C_+<C_-$ it
is accepted when $T\leq Ex_3$ and $C_+\leq Ex_4$. For each group, let $N_a$ be the number of $a$
slots and $N_b$ the number of $b$ and $b_S$ slots, and set $N=\max(N_a,N_b)$. The default thresholds
are $\{Ex_i\}=\{N+3,4,4,4\}$. ISP indices are not counted in $N$. Each item of an explicit four-tuple may be an integer or
\texttt{Automatic}, in which case only that item takes its default value. The value \texttt{None}
restores the uncut point domain. The returned \texttt{seedEnvelopeCut} metadata records the
resolved thresholds, the numbers of accepted and skipped points, and the groups emptied by the
cut.

\paragraph{Canonical-support cutting.}
The option \texttt{TotalPowerLimit} is an independent filter on the canonical support of the
generated equations:
\begin{verbatim}
ibp3 = DSGenerateIBP[allSeeds, {-1, 1}, TotalPowerLimit->7];
ibp4 = DSGenerateIBP[allSeeds, {-1, 1},
   SeedEnvelopeCut->{10, 4, 5, 4}, TotalPowerLimit->7];
\end{verbatim}
For an integer $m$, the package first combines equal integrals and removes zero coefficients. It
then retains an equation only if every canonical integral in its actual support lies in the
per-slot target box and has continuous-index absolute sum at most $m$. The default \texttt{None}
disables this filter. The two filters can be applied independently or together.

\paragraph{Generation fast path.}
In full mode, the package decomposes Association-based sealed templates into their raw integral data,
integer shifts and coefficient pieces, and memoizes canonical results by raw integral. Raw
expression lists and templates whose decomposition fails use the original symbolic path. The fast
path changes the running time, not the generated relations. On the massive-bubble benchmark it
shortens the generation stage by about a factor of three, and the relations are identical to those
of the original path.

These filters bound the generated system but do not prove closure for an arbitrary target or
basis. The flag \texttt{completeSystemQ} states completeness within the chosen envelope and
cutting policy. A new target, basis or cut setting requires an external reduction check that
every requested target reduces to the requested masters. Reproducing an uncut system requires
\texttt{SeedEnvelopeCut->None} together with \texttt{TotalPowerLimit->None}.

\subsection{Linear systems and external reduction}
\label{sec:linear}

The command
\begin{verbatim}
linearData = DSLinear[ibpData, context];
\end{verbatim}
converts the generated IBP relations into a backend-neutral association.
Its fields include the equations \texttt{linearEquations}, the complete integral
list \texttt{integralList}, the integral-to-ID mapping \texttt{integralRules}, and sector metadata.
\texttt{DSLinear} accepts the dedicated options \texttt{LinearSystemMode} (\texttt{"symbolic"} by
default; \texttt{"numeric"} builds the sampled system with the numeric rules substituted),
\texttt{CoefficientRules} (\texttt{Automatic} by default; the numeric substitutions of the sampled
mode) and \texttt{KiraOrdering} (\texttt{Automatic} by default; an association that adjusts the
integral-ordering conventions of the serializer).
The command \texttt{DSReorderIntegrals} changes this ordering before export. Master integrals
are identified at the subsequent reduction stage. The two-stage Kira plan is built in memory:
\begin{samepage}
\begin{verbatim}
prePlan   = DSKiraPlan[linearData, <|
   "stage" -> "preReduction", "candidateIntegrals" -> candidates,
   "outputDirectory" -> preDir, "jobOptions" -> Automatic|>];
preExport = DSKiraExport[prePlan];
(* ... external Kira run identifies the master integrals ... *)
linearWithUserMI = DSUserMI[linearData, userExpressions, <|
   "names" -> activeNames, "activeIndices" -> activeIndices,
   "derivativeVariables" -> {ss11, P0}, "scalingDegrees" -> degrees|>];
formalPlan   = DSKiraPlan[linearWithUserMI, <|
   "stage" -> "formal", "activeBasis" -> Automatic,
   "numericStage" -> "symbolic"|>];
formalExport = DSKiraExport[formalPlan];
(* ... external Kira run reduces to the user basis ... *)
importedData = DSKiraImport[formalWorkspace];
\end{verbatim}
\end{samepage}
The specification of \texttt{DSKiraPlan} must contain \texttt{"stage"} and accepts the further keys
\begin{itemize}
\item \texttt{"preferredIntegrals"}: integrals recorded as preferred candidates;
\item \texttt{"candidateIntegrals"}: the pre-reduction targets, the complete integral list by default;
\item \texttt{"activeBasis"}: the basis attached by \texttt{DSUserMI} when left \texttt{Automatic};
\item \texttt{"numericStage"}: \texttt{"symbolic"} by default, \texttt{"postDerivative"}
substitutes the numeric rules only after the analytic derivative closure is fixed;
\item \texttt{"coefficientRules"} and \texttt{"outputDirectory"}: forwarded to the export;
\item \texttt{"jobOptions"}: the Kira job configuration, for example the parallel job count.
\end{itemize}
The pre-reduction stage fixes the candidate
targets and lets the external reducer discover the
masters. The command \texttt{DSUserMI} then attaches the user basis. Its specification keys
\texttt{"names"} (defaulting to \texttt{"active1"}, \texttt{"active2"} and so on),
\texttt{"activeIndices"} (defaulting to all positions), \texttt{"derivativeVariables"} and
\texttt{"scalingDegrees"} (both defaulting to \texttt{Automatic}) are optional. It checks the global
and the active rank, and it stores reversible pivot rules. The formal stage constructs the analytic
derivative closure of the active basis \emph{before} any export.
\texttt{DSKiraExport} writes the input directories of a plan, or of plain \texttt{linearData}; its
options are
\begin{itemize}
\item \texttt{OutputDirectory} (\texttt{None} by default);
\item \texttt{KiraTargetIntegrals} (\texttt{Automatic}): the mandatory targets of the run;
\item \texttt{KiraCoefficientRules} (\texttt{\{\}}): the numeric coefficient rules;
\item \texttt{KiraJobOptions} (\texttt{Automatic}): the job configuration;
\item \texttt{KiraActiveBasis} (\texttt{None}): the active basis of a formal export;
\item \texttt{KiraRequireCompleteSystem} (\texttt{True}): refuse an incomplete system;
\item \texttt{KiraNumericStage} (\texttt{"symbolic"}).
\end{itemize}
A plan supplies the corresponding values from its fields.
\texttt{DSKiraImport} checks the export identity when importing the reduction. Its dedicated
options \texttt{KiraReductionFile} and \texttt{KiraMasterFile} (\texttt{Automatic} by default)
resolve to the Kira family outputs of the workspace, and \texttt{KiraCompletionFile} and
\texttt{KiraCompletionPatterns} (\texttt{Automatic} by default) control the completion diagnostics
of the run log. Kira accepts only
real rational coefficients, so the serializer rewrites the momenta that must lie on the imaginary
axis by the backend rule \texttt{k -> -I ik}, where \texttt{I} is the imaginary unit and \texttt{ik}
is the backend symbol for the same magnitude. The importer reverses this map to restore the
physical variables and their derivative Jacobians. The current export/import interface is for Kira.
The backend-neutral \texttt{linearData} representation also provides the input needed to develop
serializers for other reduction programs.

\subsection{Differential equations and checks}
\label{sec:de}

After the reduction, \texttt{DSDE} differentiates the master integrals with respect to the
independent variables:
\begin{verbatim}
deData = DSDE[importedData, {ss11, P0}];
\end{verbatim}
The variable list defaults to all initialized variables. The option
\texttt{OutputDirectory} (\texttt{None} by default) writes the DE data to disk.
Loop-Gram coordinates are handled through external-vector derivative
decompositions. Non-loop momentum magnitudes are differentiated through radial derivatives that are local to
each propagator, including
their dependent bindings. Vertex energies enter through scalar phase factors and may themselves be
included among the derivative variables. Explicit coefficients are
differentiated directly, while bare integrals are shifted and canonicalized. The result is reduced
back to the chosen active basis. The pure massive bubble example supplies a known d $\log$ basis
from \cite{Chen:2026dqp}. Constructing the DE in that basis does not require a general
algorithm for finding a d $\log$ transformation.

Under a common rescaling of momentum magnitudes and vertex energies, each integral has a definite
degree. Let $x_1,\dots,x_m$ denote these dimension-one coordinates. The spatial dimension and the
mass parameters $\nu_e$ are held fixed under this rescaling; they may enter the scaling degrees.
For a basis of homogeneous master integrals, each $I_\alpha$ satisfies the scaling
relation
\begin{equation}
\sum_{j=1}^{m} x_j\,\frac{\partial I_\alpha}{\partial x_j} \;=\; d_\alpha\, I_\alpha\,,
\label{eq:euler}
\end{equation}
where $d_\alpha$ is the constant scaling degree of $I_\alpha$. In matrix form, with
$\partial I/\partial x_j=\Omega_j(x)\,I$ for the vector of masters, the scaling relation \eqref{eq:euler}
becomes
\begin{equation}
\sum_{j=1}^{m} x_j\, \Omega_j(x) \;=\; \Delta\,,\qquad
\Delta=\mathrm{diag}(d_1,\dots,d_R)\,,
\label{eq:eulerMatrix}
\end{equation}
where $R$ is the number of masters in the active basis.
The command \texttt{DSScaleCheck} checks this matrix identity and any inhomogeneous source
terms in the DE output. It takes the DE data and an optional specification whose keys
\texttt{"relation"}, \texttt{"variables"}, \texttt{"weights"} and \texttt{"degrees"} are equivalent
to its dedicated options \texttt{ScalingRelation} (\texttt{"Custom"} by default; the built-in
relations \texttt{"PureMassiveBubble"} and \texttt{"LoopTopology"} supply the degrees of those
families when none are declared), \texttt{ScalingVariables} (\texttt{Automatic}, the DE variables),
\texttt{ScalingWeights} (\texttt{Automatic}, all weights one) and \texttt{ScalingDegrees}
(\texttt{Automatic}, the degrees declared with the active basis).
It reports a symbolic check when the scaling variables remain symbolic,
a fixed-section check when some are fixed, and an exact-point check when all are fixed.
Additional source-level time and momentum IBP checks are documented in the validation scripts;
they are separate calculations and are not part of the public interface.

At the atomic level the package also exposes the generators themselves:
\begin{equation}
\begin{aligned}
&\mathtt{dtau}[\,i,\ \mathrm{expr},\ \mathrm{ctx}\,],
&&\text{time generator at vertex } i,\\
&\mathtt{dqq}[\,i,\ j,\ \mathrm{expr},\ \mathrm{ctx}\,],
&&\frac{\partial}{\partial\, q_i\cdot q_j}\ \mathrm{expr},\\
&\mathtt{dqk}[\,i,\ j,\ \mathrm{expr},\ \mathrm{ctx}\,],
&&\frac{\partial}{\partial\, q_i\cdot k_j}\ \mathrm{expr},\\
&\mathtt{ds}[\,\mathrm{expr},\ \mathrm{var},\ \mathrm{ctx}\,],
&&\frac{d}{d\,\mathrm{var}}\ \mathrm{expr}\quad(\text{total derivative}).
\end{aligned}
\end{equation}
together with the form rewriters \texttt{rep2innerform}, \texttt{rep2outform} and
\texttt{rep2Integrand}. In these four calls the first argument of \texttt{dtau} is a vertex index,
and the first two arguments of \texttt{dqq} are loop-momentum indices, while those of \texttt{dqk}
are a loop-momentum index and an external-momentum index. The last command expands a $J$ into a
formal integrand retaining named propagator blocks, without integrating. The argument \texttt{ctx}
is optional: it carries the parsed topology association or the context established by
\texttt{DSInit}, and when it is omitted the default context set by the most recent \texttt{DSInit}
is used; explicit contexts are required once several topologies are registered.

\subsection{Tree-level IBP checks}
\label{sec:tree}

\begin{figure}[t]
\centering
\includegraphics[width=0.98\textwidth]{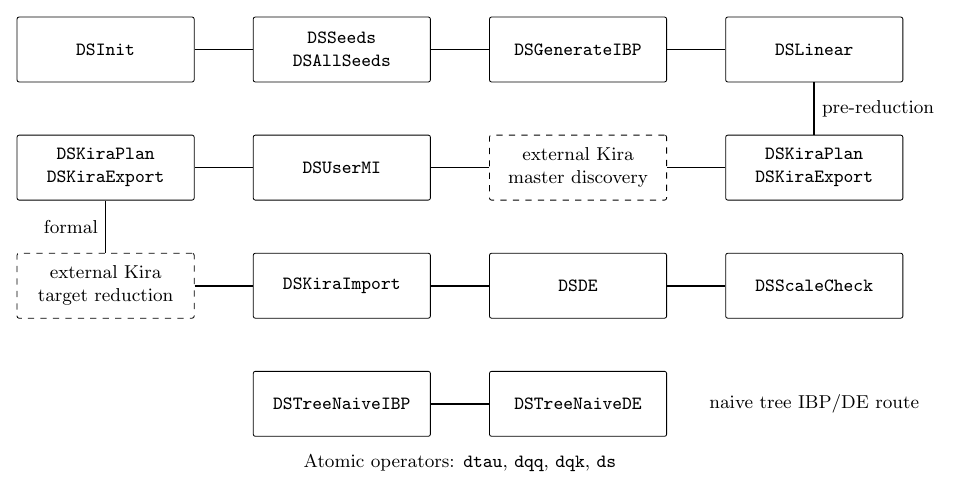}
\caption{The public workflow of \texttt{dSIBP}. Loop families in the full mode run from \texttt{DSInit}
through the seed, index-range and linear-system stages to the two-stage Kira export and import and
to the differential-equation layer. The command \texttt{DSUserMI} attaches the user basis between
the two reduction stages. The lower row shows the naive tree IBP/DE route available for
\texttt{timeOnly} families. The atomic operators are listed below the workflows.}
\label{fig:workflow}
\end{figure}

The \texttt{timeOnly} representation supports naive time-IBP checks through
\begin{verbatim}
naiveIBP  = DSTreeNaiveIBP[context, masters];
naiveDE   = DSTreeNaiveDE[naiveIBP, variables];
\end{verbatim}
The \texttt{masters} argument of \texttt{DSTreeNaiveIBP} lists the chosen tree masters, and the
\texttt{variables} argument of \texttt{DSTreeNaiveDE} defaults to the vertex energies and
massive-leg magnitudes of the family.
\texttt{DSTreeNaiveDE} also accepts the result of \texttt{DSTreeNaiveIBP} directly.
These commands construct a naive time-IBP reduction and naive tree differential equations.
Figure~\ref{fig:workflow} summarizes the loop workflow and the naive tree-level route;
table~\ref{tab:api} lists the public functions.

\section{Examples}
\label{sec:examples}
\begin{figure}[t]
\centering
\includegraphics[width=0.97\textwidth]{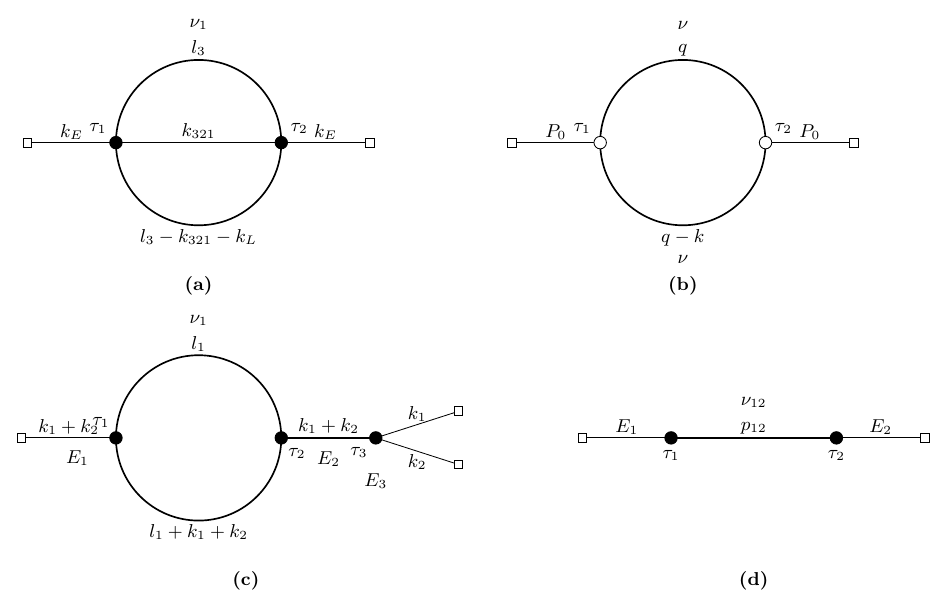}
\caption{Example diagrams. The fill color of a vertex encodes its contour sign: black is $+$ and
white is $-$. (a)~The two-loop sunrise of example~03: one massive time-ordered propagator with mass
parameter $\nu_1$ and two massless time-ordered propagators between two $+$ vertices; both vertices declare
the same vertex energy, written \texttt{kE}. (b)~The pure massive bubble of example~04: two massive
time-ordered propagators with the same mass parameter $\nu$ between two $-$ vertices with vertex energy
$P_0$; the loop external momentum is $k$. (c)~The mixed bubble-plus-tree case of example~06: a
massive time-ordered propagator with mass parameter $\nu_1$ and a massless time-ordered propagator form the loop
between $v_1$ and $v_2$. A massless tree propagator, a bridge, connects $v_2$ and $v_3$, and the
external legs carry $k_1+k_2$, $k_1$ and $k_2$. (d)~The
two-vertex tree of example~05: a massive propagator with mass parameter $\nu_{12}$ whose momentum
$p_{12}$ is a non-loop momentum magnitude.}
\label{fig:examples}
\end{figure}

The package includes six worked examples. This section first lists them together with the stage
that each one reaches, and then gives two detailed workflows. The mixed bubble-plus-tree example
shows the public commands from topology input through export for a fixed-target reduction. The pure massive bubble
example continues through a $19$-dimensional d $\log$ differential equation and its scaling check.
Figure~\ref{fig:examples} shows the diagrams. The reductions reported in these workflows were
performed with Kira; Kira's own commands are outside the scope of this section.
The distributed materials include the 19-master DE and scaling results for example~04 and the
84-master reduction summary for example~06.

\subsection{List of examples}
\label{sec:examples-list}
Each example is a Wolfram Language script, and each ends with a summary gate: it exits zero only
when every checked result matches its expectation.
\begin{itemize}
  \item \texttt{01\_mixed\_bubble\_workflow.wl}: a bubble with one massive and one massless time-ordered
        propagator; runs the unified public entry through the discrete templates, the continuous
        sweep and the linear system, and builds a pre-reduction plan in memory;
  \item \texttt{02\_function\_system\_hankel.wl}: a bare Hankel function system $(P,Q,T,W)$ in
        the \texttt{timeOnly} mode; initialization and seed generation through the custom
        function-system interface;
  \item \texttt{03\_single\_massive\_sunrise}: the two-loop single-mass sunrise of
        figure~\ref{fig:examples}a; a complete set of general IBP seeds and parameter-derivative operators,
        without a continuous index sweep or reduction;
  \item \texttt{04\_pure\_massive\_bubble\_closed\_loop}: the pure massive bubble of
        figure~\ref{fig:examples}b; seeds, index sweep, linear system, a user-supplied active
        basis, reduction, d $\log$ differential equations and a symbolic Euler scaling check;
  \item \texttt{05\_tree\_two\_vertex\_time\_ibp}: the massive two-vertex tree of
        figure~\ref{fig:examples}d in the \texttt{timeOnly} mode, in which the IBP relations are
        generated computationally from total time derivatives; the example also exercises
        generic \texttt{timeOnly} seeds, \texttt{DSGenerateIBP}, \texttt{linearData}, the
        naive tree reduction and DE, and the massless \texttt{PendingRederivation} boundary;
  \item \texttt{06\_mix\_bubble\_tree}: the mixed bubble-plus-tree case of
        figure~\ref{fig:examples}c; a complete set of IBP seeds and saved summaries of an uncut continuous
        sweep, Kira export and a fixed-target reduction.
\end{itemize}
Example 01 reaches a sampled linear system that can be exported to Kira. Example 03 stops at
general symbolic seeds and does not produce \texttt{linearData}. For tree families without loop
momenta, MadStree \cite{Chen:2026iwf} provides the reduction and d $\log$ differential equations.

\subsection{Example: pure massive bubble through \texorpdfstring{d $\log$}{d log} DE and scaling}
\label{sec:example-bubble}

\subsubsection{Family definition and IBP system}

The pure massive bubble has two $-$ vertices of energy $P_0$ and two massive time-ordered propagators with
the same $\nu$, carrying $q$ and $q-k$. Its active coordinates are $\texttt{ss}11=|k|$ and $P_0$.
The input uses $a_0[v_i]=2\nu+\texttt{analyticRegulator}$ and $b_0[e]=-2\nu$. The released example
uses the frozen Negative convention to reproduce its reference basis. Its family declaration and
core initialization settings are
\begin{verbatim}
caseInput = <|
   "name" -> "pureMassiveBubbleMinusMinus",
   "vertices" -> {
     <|"id" -> v1, "vertexType" -> "-", "externalLegEnergy" -> P0|>,
     <|"id" -> v2, "vertexType" -> "-", "externalLegEnergy" -> P0|>
     },
   "lines" -> {
     <|"id" -> 1, "endpoints" -> {v1, v2}, "momentum" -> q,
       "massType" -> "massive", "nu" -> nu|>,
     <|"id" -> 2, "endpoints" -> {v1, v2}, "momentum" -> q - k,
       "massType" -> "massive", "nu" -> nu|>
     },
   "loopMomenta" -> {q},
   "loopExternalMomenta" -> {k},
   "independentExternalMomenta" -> {},
   "ibpMode" -> "full",
   "ispData" -> {},
   "zeroPointRules" -> {
     a0[v1] -> 2 nu + analyticRegulator,
     a0[v2] -> 2 nu + analyticRegulator,
     b0[1] -> -2 nu, b0[2] -> -2 nu
     },
   "symmetryRules" -> exampleSymmetryRules0
   |>;

context = DSInit[
   caseInput,
   NuConvention -> "Negative",
   WriteInitializationFiles -> True,
   InitializationDirectory -> FileNameJoin[{exampleWorkspace, "init"}],
   GenerateDerivativeMetadata -> True,
   OverwriteInitialization -> True
   ];
\end{verbatim}
Initialization metadata are written under \texttt{exampleWorkspace}, outside the repository.

The distributed \texttt{main.wl} sets this workspace through \texttt{DSIBP\_KIRA\_WORKSPACE}.

The helper \texttt{exampleSymmetryRules0} implements the reference parity and canonical vertex
and propagator exchanges. The target envelope \texttt{referenceTopTargetEnvelope} is also supplied
with the example. With this context, the general seeds and the continuous system are constructed by
\begin{verbatim}
seedData = DSSeeds[context];
allSeeds = DSAllSeeds[seedData];
generatedIBP = DSGenerateIBP[allSeeds,
   Sequence @@ referenceTopTargetEnvelope];
linearData = DSLinear[generatedIBP, context,
   LinearSystemMode -> "numeric",
   CoefficientRules -> parameterProbeRules,
   KiraOrdering -> Automatic];
\end{verbatim}
The target envelope is $a_{1,2}\in[0,3]$ and $b_{1,2}\in[-3,5]$, with
\texttt{SeedEnvelopeCut -> Automatic} in this example.

\subsubsection{General-index derivatives and EOM canonicalization}

For
$J_0=J[\{a[v_1],a[v_2]\},\{\{b[1],0,0\},\{b[2],0,0\}\},\{\}]$, the vertex-energy derivative is
simply
\begin{equation}
\frac{\partial J_0}{\partial P_0}
=-\mathrm i\,J_0\big|_{a[v_2]\to a[v_2]+1}
-\mathrm i\,J_0\big|_{a[v_1]\to a[v_1]+1}.
\label{eq:bubbleP0Derivative}
\end{equation}
It is obtained by \texttt{ds[J0, P0, topology]}. The call
\texttt{ds[J0, ss11, topology]} additionally shifts the propagator powers and the massive $n$
states, including $b_e\to b_e\pm2$ and $a_v\to a_v+1$.

These derivatives enter the same \texttt{DSSeeds} canonicalization chain described above,
including EOM and endpoint rules, before the continuous sweep.

\subsubsection{The active \texorpdfstring{d $\log$}{d log} basis}

The reference basis is inserted as 19 user-defined linear combinations, rather than asking the
backend to discover a d $\log$ basis:
\begin{verbatim}
linearDataWithUserMI = DSUserMI[linearData,
   pureMassiveBubbleUserMIExpressions, <|
    "names" -> pureMassiveBubbleUserMINames,
    "activeIndices" -> Range[19],
    "derivativeVariables" -> {ss11, P0},
    "scalingDegrees" -> pureMassiveBubbleUserMIScalingDegrees|>];
formalPlan = DSKiraPlan[linearDataWithUserMI, <|
   "stage" -> "formal", "activeBasis" -> Automatic,
   "numericStage" -> "postDerivative",
   "coefficientRules" -> parameterProbeRules,
   "outputDirectory" -> kiraDir, "jobOptions" -> kiraJobOptions|>];
DSKiraExport[formalPlan];
\end{verbatim}
For completeness, the 19 active combinations are listed below in the package's $J$ notation. The
first five are long linear combinations; the remaining fourteen are shorter monomials and
pinched-sector terms.
\begin{verbatim}
B1 = 4 J[{0,0},{{0,0,0},{3,0,1}},{}]
   - 4 nu J[{0,0},{{1,0,1},{2,0,0}},{}]
   + 2 J[{0,1},{{0,0,0},{2,0,0}},{}]
   - 2 J[{0,1},{{0,0,0},{2,1,1}},{}]
   - 2 J[{0,1},{{1,0,1},{1,0,1}},{}]
   - 2 J[{0,1},{{1,0,1},{1,1,0}},{}]

B2 = 2 J[{0,1},{{-1,0,1},{3,1,0}},{}]
   + 2 J[{0,1},{{-1,1,0},{3,0,1}},{}]
   + (4+4 nu) J[{0,1},{{0,0,0},{2,1,1}},{}]
   - 4 nu J[{0,1},{{0,1,1},{2,0,0}},{}]
   - 2 ss11^2 J[{0,1},{{1,0,1},{3,1,0}},{}]
   - 2 ss11^2 J[{0,1},{{1,1,0},{3,0,1}},{}]
   - J[{0,2},{{-1,0,1},{2,1,1}},{}]
   + J[{0,2},{{-1,1,0},{2,0,0}},{}]
   + 3 J[{0,2},{{0,0,0},{1,1,0}},{}]
   - 3 J[{0,2},{{0,1,1},{1,0,1}},{}]
   + ss11^2 J[{0,2},{{1,0,1},{2,1,1}},{}]
   - ss11^2 J[{0,2},{{1,1,0},{2,0,0}},{}]
   + J[{1,1},{{-1,0,1},{2,0,0}},{}]
   - J[{1,1},{{-1,0,1},{2,1,1}},{}]
   + 3 J[{1,1},{{0,0,0},{1,0,1}},{}]
   - 3 J[{1,1},{{0,1,1},{1,0,1}},{}]
   - ss11^2 J[{1,1},{{1,0,1},{2,0,0}},{}]
   + ss11^2 J[{1,1},{{1,0,1},{2,1,1}},{}]

B3 = -4 J[{0,1},{{-1,0,1},{3,0,1}},{}]
   - 8 nu J[{0,1},{{0,0,0},{2,0,0}},{}]
   + 4 ss11^2 J[{0,1},{{1,0,1},{3,0,1}},{}]
   - 2 J[{0,2},{{-1,0,1},{2,0,0}},{}]
   - 6 J[{0,2},{{0,0,0},{1,0,1}},{}]
   + 2 ss11^2 J[{0,2},{{1,0,1},{2,0,0}},{}]
   + 2 J[{1,1},{{-1,0,1},{2,1,1}},{}]
   - 4 J[{1,1},{{0,0,0},{1,0,1}},{}]
   + 2 J[{1,1},{{0,1,1},{1,0,1}},{}]
   - 2 ss11^2 J[{1,1},{{1,0,1},{2,1,1}},{}]

B4 = 4 J[{0,0},{{0,1,1},{3,0,1}},{}]
   + 4 J[{0,0},{{1,0,1},{2,1,1}},{}]
   + 4 nu J[{0,0},{{1,0,1},{2,1,1}},{}]
   + 2 J[{0,1},{{0,1,1},{2,0,0}},{}]
   - 2 J[{0,1},{{0,1,1},{2,1,1}},{}]
   + 2 J[{0,1},{{1,0,1},{1,1,0}},{}]
   + 2 J[{0,1},{{1,1,0},{1,1,0}},{}]

B5 = -4 J[{0,1},{{-1,1,0},{3,1,0}},{}]
   + (8+8 nu) J[{0,1},{{0,1,1},{2,1,1}},{}]
   + 4 ss11^2 J[{0,1},{{1,1,0},{3,1,0}},{}]
   + 2 J[{0,2},{{-1,1,0},{2,1,1}},{}]
   + 6 J[{0,2},{{0,1,1},{1,1,0}},{}]
   - 2 ss11^2 J[{0,2},{{1,1,0},{2,1,1}},{}]
   - 2 J[{1,1},{{-1,0,1},{2,0,0}},{}]
   - 2 J[{1,1},{{0,0,0},{1,0,1}},{}]
   + 4 J[{1,1},{{0,1,1},{1,0,1}},{}]
   + 2 ss11^2 J[{1,1},{{1,0,1},{2,0,0}},{}]

B6  = ss11 J[{0,0},{{2,0,0},{2,0,0}},{}]
B7  = ss11 J[{0,0},{{2,0,0},{2,1,1}},{}]
B8  = ss11 J[{0,0},{{2,1,1},{2,1,1}},{}]
B9  = ss11 J[{0,1},{{1,0,1},{2,0,0}},{}]
B10 = ss11 J[{0,1},{{1,0,1},{2,1,1}},{}]
B11 = ss11 J[{0,1},{{1,1,0},{2,0,0}},{}]
B12 = ss11 J[{0,1},{{1,1,0},{2,1,1}},{}]
B13 = 2 ss11 J[{0,0},{{1,0,1},{3,0,1}},{}]
B14 = 2 ss11 J[{0,0},{{1,0,1},{3,1,0}},{}]
B15 = J[{0},{{2},{2,0,0}},{}]
B16 = J[{0},{{2},{2,1,1}},{}]
B17 = -(2 I/P0) (1+2 nu) J[{0},{{2},{1,0,1}},{}]
B18 = J[{1},{{2},{1,0,1}},{}]
B19 = (4 J[{0},{{0},{3,0,1}},{}]
     + 2 J[{1},{{0},{2,0,0}},{}]
     - 2 J[{1},{{0},{2,1,1}},{}])/ss11
\end{verbatim}
All 19 entries are active; there are no auxiliary user-MI relations. Their common scaling degree is
$\dim-5$.

\subsubsection{Retrieved reduction, DE and scaling}

After the external reduction has returned \texttt{kira\_list.m}, \texttt{masters} and the Kira log,
the package reconstructs the physical coefficients and builds the DE:
\begin{verbatim}
reductionData = DSKiraImport[kiraDir, context];
deData = DSDE[reductionData, {ss11, P0},
   OutputDirectory -> deDir];
scaleData = DSScaleCheck[deData, <|
   "relation" -> "PureMassiveBubble",
   "variables" -> {ss11, P0}, "weights" -> {1, 1},
   "degrees" -> (pureMassiveBubbleUserMIScalingDegrees /.
      parameterProbeRules)|>];
\end{verbatim}
The run produces two $19\times19$ matrices in $\texttt{ss}11$ and $P_0$, with no residual $J$
or backend token. The Euler matrix and source relations are both satisfied symbolically. The matrices
also agree entry by entry with the independent reference of \cite{Chen:2026dqp} in the
$\texttt{ss}11$, $P_0$ and \texttt{ip0} coordinates.

\subsection{Example: mixed bubble with a tree bridge}
\label{sec:example-mix}

The diagram is shown in figure~\ref{fig:examples}c. Propagators~1 and~2 form the loop:
propagator~1 is a massive time-ordered propagator with mass parameter $\nu_1$ carrying $l_1$, and
propagator~2 is a massless time-ordered propagator carrying $l_1+k_1+k_2$. Propagator~3 is a massless tree
bridge between $v_2$ and $v_3$. It lies outside the loop, so its pack carries no propagator power:
the first slot is the placeholder \texttt{"F"}, and the momentum-magnitude power of $k_1+k_2$ is
collected in the sector normalization $N_s$. One external leg enters $v_1$ with momentum $k_1+k_2$,
and two external legs enter $v_3$ with momenta $k_1$ and $k_2$. The three vertex energies are
$E_1,E_2,E_3$.

\begin{figure}[t]
\centering
\includegraphics[width=0.62\textwidth]{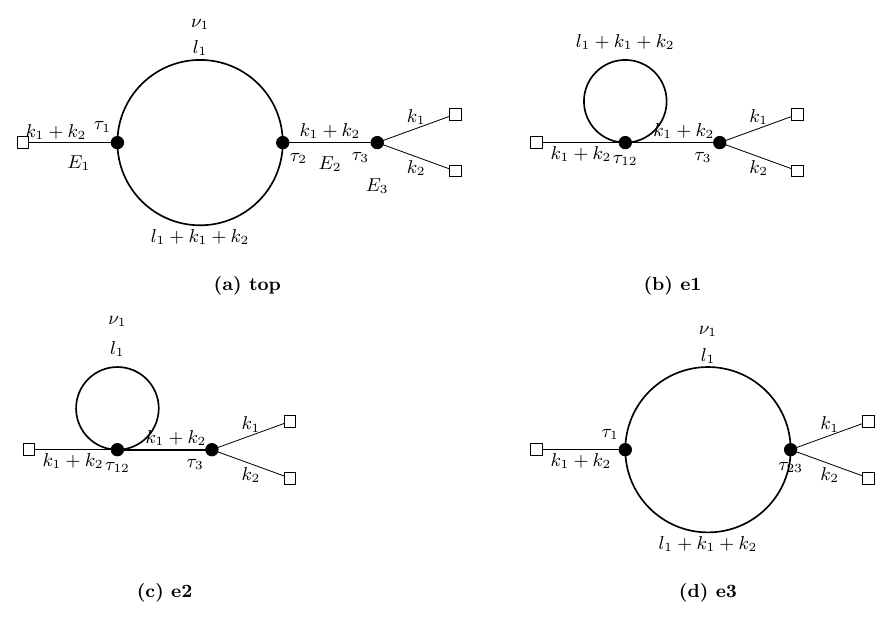}
\caption{The mixed bubble-plus-tree example and its pinched subsectors. (a)~The top sector.
(b)~Pinch of the massive cycle propagator before the massless-tadpole fold. (c)~Pinch
of the massless cycle propagator. (d)~Pinch of the bridge. Vertex fills and propagator labels
are as in figure~\ref{fig:examples}.}
\label{fig:examples06}
\end{figure}

The six sectors and their normalizations are illustrated below with selected masters.

\subsubsection{Input and initialization}

The complete physical declaration is
\begin{verbatim}
caseInput = <|
   "name" -> "mixBubbleTreeK1K2",
   "vertices" -> {
     <|"id" -> v1, "vertexType" -> "+", "externalLegEnergy" -> E1|>,
     <|"id" -> v2, "vertexType" -> "+", "externalLegEnergy" -> E2|>,
     <|"id" -> v3, "vertexType" -> "+", "externalLegEnergy" -> E3|>},
   "lines" -> {
     <|"id" -> 1, "endpoints" -> {v1, v2}, "momentum" -> l1,
       "nu" -> nu1, "massType" -> "massive"|>,
     <|"id" -> 2, "endpoints" -> {v1, v2}, "momentum" -> l1+k1+k2,
       "massType" -> "massless"|>,
     <|"id" -> 3, "endpoints" -> {v2, v3}, "momentum" -> k1+k2,
       "massType" -> "massless"|>},
   "extLegs" -> {{bubbleLeg, v1, k1+k2},
     {treeLeg1, v3, k1}, {treeLeg2, v3, k2}},
   "loopMomenta" -> {l1},
   "loopExternalMomenta" -> {k1+k2},
   "independentExternalMomenta" -> {k1, k2},
   "ibpMode" -> "full", "ispData" -> {},
   "zeroPointRules" -> {
     a0[v1] -> alpha1, a0[v2] -> alpha2, a0[v3] -> alpha3,
     b0[1] -> beta1, b0[2] -> beta2, b0[3] -> beta3},
   "parityConstraints" -> {
     b[1]+n[1,1]+n[1,2] -> 0, b[2]+n[2,1]+n[2,2] -> 0}|>;
\end{verbatim}
The two parity constraints select the even-even (EE) subsystem of the two cycle propagators. The
bridge is not a loop propagator and is excluded from them.

Initialization compiles the topology, propagator systems, sectors and derivative metadata. We keep
the default-coordinate context for the operator examples below, and a second, redefined context for
the reduction:
\begin{verbatim}
context0 = DSInit[caseInput, NuConvention -> "Positive",
   GenerateDerivativeMetadata -> True, RegisterAsCurrent -> True,
   WriteInitializationFiles -> False];
topology = Lookup[context0, "topology", <||>];
context = DSRedefineParameters[context0, {
   sp[k1+k2,k1+k2] -> loopScale^2,
   sp[k1,k1] -> legScale1^2, sp[k2,k2] -> legScale2^2}];
customTopology = Lookup[context, "topology", <||>];
\end{verbatim}
The derivative variables of the redefined context are $\texttt{loopScale}$,
$\texttt{legScale1}$, $\texttt{legScale2}$, $E_1$, $E_2$ and $E_3$.

\subsubsection{General seeds and canonicalization}

The public operators \texttt{dtau}, \texttt{dqq} and \texttt{dqk} construct time, loop-momentum and
external-momentum derivatives directly in the three-slot $J$ representation. For example,
\begin{verbatim}
integral = J[{0,0,0},
   {{0,0,0},{0,0,0},{"F",0,1}}, {}];
timeSeed = dtau[v3, integral, topology];
loopLoopSeed = dqq[1, 1, integral, topology];
\end{verbatim}
The time derivative gives the compact four-term relation
\begin{verbatim}
-2 J[{0,0}, {{0,0,0},{0,0,0},{"F"}}, {}]
+ alpha3 J[{0,0,-1}, {{0,0,0},{0,0,0},{"F",1,0}}, {}]
+ I ss11 J[{0,0,0}, {{0,0,0},{0,0,0},{"F",0,0}}, {}]
+ I E3 J[{0,0,0}, {{0,0,0},{0,0,0},{"F",1,0}}, {}].
\end{verbatim}
The corresponding loop-momentum seed is
\begin{verbatim}
(beta2/2) J[{0,0,0},{{-2,0,0},{2,0,0},{"F",1,0}},{}]
+ (beta1+beta2/2-dim) J[{0,0,0},
  {{0,0,0},{0,0,0},{"F",1,0}},{}]
- (I/2) J[{0,1,0},{{-2,0,0},{1,1,0},{"F",1,0}},{}]
- J[{0,1,0},{{-1,0,1},{0,0,0},{"F",1,0}},{}]
- (I/2) J[{0,1,0},{{0,0,0},{-1,1,0},{"F",1,0}},{}]
+ (I/2) J[{1,0,0},{{-2,0,0},{1,1,0},{"F",1,0}},{}]
- J[{1,0,0},{{-1,1,0},{0,0,0},{"F",1,0}},{}]
+ (I/2) J[{1,0,0},{{0,0,0},{-1,1,0},{"F",1,0}},{}]
- (beta2 kk[1,1]/2) J[{0,0,0},
  {{0,0,0},{2,0,0},{"F",1,0}},{}]
+ (I kk[1,1]/2) J[{0,1,0},
  {{0,0,0},{1,1,0},{"F",1,0}},{}]
- (I kk[1,1]/2) J[{1,0,0},
  {{0,0,0},{1,1,0},{"F",1,0}},{}].
\end{verbatim}
In the four-term time seed, the first term contains the pinched bridge, while the remaining
three retain an unpinched bridge block; the third and fourth terms carry the state $(1,0)$.
The loop-momentum seed contains the corresponding shifts of $a$, $b$ and the two massive $n$
slots, including the pinched shifts $b_e\to b_e\pm2$.

These atomic expressions are not yet the final complete set of IBP seeds. \texttt{DSSeeds} enumerates the
discrete $n=0,1$ states and applies the massive EOM recursion, massless endpoint canonicalization,
parity and sector rules. Only then does \texttt{DSGenerateIBP} expand the continuous indices:
\begin{verbatim}
seedData = DSSeeds[context, ProgressReporting -> False];
allSeeds = DSAllSeeds[seedData];
generatedIBP = DSGenerateIBP[allSeeds,
   {a[v1],0,5}, {a[v2],0,5}, {a[v3],-1,1},
   {b[1],-4,7}, {b[2],-4,7},
   SeedEnvelopeCut -> None, TotalPowerLimit -> None,
   ProgressReporting -> False];
\end{verbatim}
These ranges give the final stable root envelope. The package maps it to each subsector and
then derives the actual seed range from the shifts of every template.

\subsubsection{Linear system, target ordering and export}

For the reduction recorded here, the continuous relations are converted to a \texttt{linearData}
association with the recorded exact rational rules, while \texttt{beta1} and \texttt{beta2} remain symbolic. The
target is the fixed set of 903 integrals with $a_3=0$, nonnegative continuous slots and
$\sum|a|+\sum|b|\leq5$. A second ordering pass places all targets before the remaining integrals
without changing their internal dSIBP order:
\begin{verbatim}
linearData = DSLinear[generatedIBP, context,
   LinearSystemMode -> "numeric",
   CoefficientRules -> exactPointRules, KiraOrdering -> Automatic];
targets = Select[linearData["integralList"], targetQ];
linearOrdered = DSLinear[generatedIBP, context,
   LinearSystemMode -> "numeric",
   CoefficientRules -> exactPointRules,
   KiraOrdering -> <|"PreferredIntegrals" -> targets,
     "PreferredPriority" -> "BeforeB"|>];
plan = DSKiraPlan[linearOrdered, <|
   "stage" -> "preReduction", "candidateIntegrals" -> targets,
   "coefficientRules" -> exactPointRules,
   "outputDirectory" -> winDir, "jobOptions" -> kiraJobOptions|>];
DSKiraExport[plan];
\end{verbatim}
Here \texttt{targetQ} implements the stated nonnegative box. The exported system has 205,434
canonical equations and 85,306 integrals. The distributed \texttt{main.wl} rechecks initialization,
the complete set of IBP seed templates and atomic relations, and reads the saved Kira input and result summaries;
it does not repeat the continuous sweep or the reduction.

The final reduction uses the root envelope $a_{1,2}\in[0,5]$, $a_3\in[-1,1]$ and
$b_{1,2}\in[-4,7]$. The fixed 903 targets reduce to 84 master integrals.

\subsubsection{Reading propagator blocks through selected masters}

With $\Delta_{23}=\tau_2-\tau_3$, $q_3=|k_1+k_2|$ and the common contour sign $\sigma=+1$, the block
carried by $\{\texttt{"F"},n_{3,1},n_{3,2}\}$ is
\begin{equation}
\begin{aligned}
\mathcal B_3^{(0,0)}&=\theta(\Delta_{23})\,e^{-\mathrm{i}q_3\Delta_{23}}
+\theta(-\Delta_{23})\,e^{+\mathrm{i}q_3\Delta_{23}},\\
\mathcal B_3^{(1,0)}&=-\theta(\Delta_{23})\,e^{-\mathrm{i}q_3\Delta_{23}}
+\theta(-\Delta_{23})\,e^{+\mathrm{i}q_3\Delta_{23}},
\end{aligned}
\label{eq:bridgeBlock}
\end{equation}
and the remaining states follow from \eqref{eq:presetWronskians} and \eqref{eq:masslessQuotient}. Because $q_3$ is a non-loop
magnitude, the pack attaches no denominator to it: the declared zero point $\beta_3$ enters through
$N_s$. By contrast, the massless cycle magnitude $\xi_2=|l_1+k_1+k_2|$ depends on the loop momentum,
so its power remains in the integrand as $\xi_2^{-(b_2+\beta_2)}$.

The family contains six sectors. After the massless tadpole is folded to a pinched propagator,
their keys are \texttt{top}, \texttt{e1\_e2}, \texttt{e2}, \texttt{e3},
\texttt{e1\_e2\_e3} and \texttt{e2\_e3}. The sector normalizations are
\begin{equation}
N_s=\begin{cases}
q_3^{-\beta_3}, & s\in\{\texttt{top},\texttt{e2},\texttt{e3},\texttt{e2\_e3}\},\\[2pt]
-\dfrac{4\mathrm{i}}{\pi}\,q_3^{-\beta_3}, &
s\in\{\texttt{e1\_e2},\texttt{e1\_e2\_e3}\},
\end{cases}
\label{eq:ex06Ns}
\end{equation}
for real $\nu_1$. Only pinching the massive cycle propagator produces the massive
pinched-term constant.

\texttt{rep2Integrand} expands a selected $J$ into phases, time powers, momentum powers and named
propagator blocks. For example,
\begin{verbatim}
rep2Integrand[J[{0,0,0},
  {{0,0,0},{0,0,0},{"F",0,0}},{}], topology]
rep2Integrand[J[{0,0,0},
  {{0,1,1},{0,0,0},{"F",1,0}},{}], topology]
\end{verbatim}
The first expression contains
$\mathcal B_1^{(0,0)}\mathcal B_2^{(0,0)}\mathcal B_3^{(0,0)}$ and the powers
$\xi_1^{-\beta_1}\xi_2^{-\beta_2}q_3^{-\beta_3}$. The second replaces the massive and bridge blocks
by $\mathcal B_1^{(1,1)}$ and $\mathcal B_3^{(1,0)}$, making the two discrete-state slots explicit.

Three pinched-sector examples show the other pack types:
\begin{itemize}
  \item $J[\{0,0\},\{\{0,0,0\},\{1,1,0\},\{\texttt{"F"}\}\},\{\}]$ is an \texttt{e3} master.
        The bridge is pinched, and the remaining massless cycle block is
        $\mathcal B_2^{(1,0)}$ with $\xi_2^{-1-\beta_2}$.
  \item $J[\{0,0\},\{\{0,1,1\},\{1\},\{\texttt{"F"},0,0\}\},\{\}]$ is an \texttt{e2} master.
        It contains a massive tadpole block $\mathcal B_1^{(1,1)}$, the pinched-cycle power
        $\xi_2^{-1-\beta_2}$ and an unpinched bridge block.
  \item $J[\{0,0\},\{\{0\},\{0\},\{\texttt{"F"},1,0\}\},\{\}]$ belongs to \texttt{e1\_e2}.
        Both cycle lines are pinched. There is no remaining cycle block; the remaining term coming from pinching a massive propagator
        supplies $-4\mathrm i/\pi$ and $\xi_1^{-\beta_1+2\nu_1}$, while the bridge still supplies
        $\mathcal B_3^{(1,0)}$.
\end{itemize}
These examples also show why the sector normalization and the integer slots must be read together:
a pinched line changes the explicit momentum power and the time zero point, not merely the pack
length.

\subsection{Validation}
\label{sec:validation}

The DE and IBP results are compared at specified points against independent calculations;
agreement there is not a symbolic proof for all kinematics.
\begin{enumerate}
\item \emph{Tree level.} For a massive two-vertex tree, the five-master IBP relations and the
corresponding DE are checked against Sec.~4 of \cite{Chen:2024glu}.
\item \emph{Loop level.} For the pure massive bubble, the nineteen-master reduction and the
$19\times19$ d $\log$ DE are checked against the supplementary matrices of \cite{Chen:2026dqp}.
The Euler scaling relation is checked for the same nineteen-master basis.
\item \emph{Derivative operators.} The \texttt{dSIBP} derivative operators are checked against the
direct d $\log$ DE construction of MadStree \cite{Chen:2026iwf}.
\end{enumerate}

\section{Summary}
\label{sec:summary}

\texttt{dSIBP} organizes the structures that arise in loop IBP for de Sitter diagrams and assigns
compact notation to them. Starting from the topology of a Feynman diagram, it constructs a
complete set of IBP seeds and derivative operators with respect to the dynamical variables, and uses externally
obtained reductions to construct differential equations, providing input for subsequent
phenomenological calculations.

Additional relations for massless time-ordered propagators are built into the package. These
include the common time-ordering merger and splitting rules for banana sub-diagrams
(appendix~\ref{app:banana}) and the extra folds of massless banana propagators
(appendix~\ref{app:fold}).

\section*{Acknowledgements}
We thank the developers of Kira \cite{Maierhofer:2017gsa,Klappert:2020nbg}, Rational Tracer
\cite{Magerya:2022hvj} and MadStree \cite{Chen:2026iwf} for making their tools available. Jiaqi Chen is supported by the National Natural Science Foundation of China (NSFC) under
Grant No.~12505094 and by the Science Foundation of China University of Petroleum, Beijing
(No.~2462025YJRC019).

\medskip
\noindent\textbf{Statement on AI Use.}
The authors developed scientific ideas and methodology and designed the function interfaces, commands, and invocation patterns of the software package.
Several agent programs driven by different large language models were used in this work. The authors have reviewed and verified all the results.

\clearpage
\appendix

\section{Banana sub-diagrams: the common time-ordering merger and splitting into remaining terms}
\label{app:banana}

This appendix collects the two formulas behind the remaining terms \eqref{eq:pinched} of a banana
sub-diagram, for an arbitrary number of members and an arbitrary mixture of massive and massless
time-ordered propagators. Write every member $e$ as two time-ordering halves,
\begin{equation}
G_e=\theta(\Delta)\,A_e+\theta(-\Delta)\,B_e,
\qquad \Delta=\tau_u-\tau_v,
\label{eq:bananaHalves}
\end{equation}
where $A_e$ is the half with $\tau_u>\tau_v$ and $B_e$ the half with $\tau_v>\tau_u$: for the
massless blocks \eqref{eq:masslessBlocks} they are the two phase factors, and for a massive block
they are the two time-ordered halves of the mode-function kernel. Because all members share the same $\Delta$, every mixed
time-ordering product is supported on $\Delta=0$ and vanishes almost everywhere, so the product
over the banana sub-diagram merges to
\begin{equation}
\prod_{e\in B}G_e
=\theta(\Delta)\prod_{e\in B}A_e+\theta(-\Delta)\prod_{e\in B}B_e,
\label{eq:bananaMerger}
\end{equation}
and its singular part under the time derivative is the single boundary
\begin{equation}
\biggl(\,\partial_\Delta\prod_{e\in B}G_e\biggr)_{\mathrm{singular}}
=\delta(\Delta)\Bigl(\prod_{e\in B}A_e-\prod_{e\in B}B_e\Bigr).
\label{eq:bananaBoundary}
\end{equation}
With $\mathsf J_e=(A_e+B_e)/2$, the equal-time symmetric factor of one member, and the jump
$D_e=A_e-B_e$ fixed by \(W_T\), which is given in \eqref{eq:presetWronskians}, for a massive member (the symbol $\mathsf J_e$ is not the integral
head $J$), the difference of the two
products splits into odd subsets,
\begin{equation}
\prod_{e\in B}A_e-\prod_{e\in B}B_e
=\sum_{\substack{\varnothing\neq S\subseteq B\\ |S|\ \mathrm{odd}}}
2^{1-|S|}\prod_{e\in S}D_e\prod_{e\notin S}\mathsf J_e\,.
\label{eq:bananaSplitting}
\end{equation}
The derivation is one line: substituting $A_e=\mathsf J_e+D_e/2$ and
$B_e=\mathsf J_e-D_e/2$ expands both products into the same monomials, the monomials with an even
number of $D$ factors cancel, and each odd subset $S$ survives with coefficient $2^{1-|S|}$. One
pinch event therefore pinches any nonempty odd subset $S$ of the banana sub-diagram with
coefficient
$2^{1-|S|}$, merges the two vertices once and carries a single $\delta$: for two members only the
single-line pinched remaining term occurs, for three members the triple-line pinched remaining term occurs with
coefficient $1/4$, and the members left
over are immediately reduced by the built-in coincident-vertex relations, so no further boundary
appears. The equal-time value $\theta(0)=\tfrac12$ enters only through $\mathsf J_e$; the package
never multiplies point values of $\delta$ against the remaining $\theta$ factors.

\section{Folds of massless banana propagators}
\label{app:fold}

This appendix derives the built-in relation for massless banana propagators in the original
index representation, without combining their momenta. Let $e_1,\dots,e_p$ be the massless time-ordered
propagators of a banana sub-diagram, ordered by line index, each carrying a
quotient state $(n_{e_k,1},0)$ with $n_{e_k,1}\in\{0,1\}$. All members share the time difference
$\Delta=\tau_u-\tau_v$ and the contour sign $\sigma$ of the pair, and each has its own momentum
magnitude $q_{e_k}$. Split each block of \eqref{eq:masslessBlocks} into its two time-ordering
parts,
\begin{equation}
A_k=\theta(\Delta)\,e^{-\mathrm{i}\sigma q_{e_k}\Delta},
\qquad
B_k=\theta(-\Delta)\,e^{+\mathrm{i}\sigma q_{e_k}\Delta},
\label{eq:foldParts}
\end{equation}
so that $M_0^{(k)}=A_k+B_k$ and $M_1^{(k)}=B_k-A_k$, and a member with state $n_k$ contributes
the factor $F_{e_k}[n_k,0]=B_k+(1-2n_k)A_k$. Since $\theta(\Delta)\theta(-\Delta)=0$, every
mixed product of two members vanishes identically, and the product over the banana sub-diagram
collapses to
\begin{equation}
\prod_{k=1}^{p}F_{e_k}[n_k,0]
=\Bigl(\prod_{k=1}^{p}B_k\Bigr)
+(-1)^{\,\sum_{k}n_k}\Bigl(\prod_{k=1}^{p}A_k\Bigr).
\label{eq:foldProduct}
\end{equation}
The kernel product therefore depends on the member states only through the total parity
$\sum_k n_k\bmod 2$: two state assignments of the same banana sub-diagram define the same
integrand whenever
their parities agree. In integral labels the weak form of the fold is the relation
\begin{equation}
J\bigl[\dots\{b_{e_k},n_{e_k},0\}\dots\bigr]
=J\bigl[\dots\{b_{e_k},n'_{e_k},0\}\dots\bigr],
\qquad
\sum_{k}n_{e_k}\equiv\sum_{k}n'_{e_k}\pmod 2\,.
\label{eq:foldWeakForm}
\end{equation}

The package canonicalizes a banana sub-diagram by moving every excitation onto its first member: the
canonical representative assigns the parity $P=\sum_k n_k\bmod2$ to $e_1$ and the state $(0,0)$
to every other member. Equation~\eqref{eq:foldProduct} shows that the move leaves the integrand
unchanged when the members are oriented consistently. If instead a member $e_k$ is declared with
reversed ordered endpoints, $\Delta\to-\Delta$ exchanges $A_k$ and $B_k$ in \eqref{eq:foldParts},
so $M_0$ is even and $M_1$ is odd under the exchange,
\begin{equation}
F_{e_k}[1,0](v,u)=-\,F_{e_k}[1,0](u,v),
\qquad
F_{e_k}[0,0](v,u)=F_{e_k}[0,0](u,v),
\label{eq:foldExchange}
\end{equation}
and moving an excitation to or from a reversed member multiplies the integrand by $(-1)^{n_{e_k}}$;
in the canonicalization this appears as one factor $-1$ per reversed member that carries $n=1$.

The fold is an algebraic canonicalization layer, applied after the single-line quotient
\eqref{eq:masslessQuotient} whenever seed templates are canonicalized, and the resulting
fold relations are recorded together with the sector data. Seed enumeration still sweeps
the one-sided states of a banana sub-diagram separately, because pinches of different
members reach different subsectors and their boundary equations carry independent
information. The canonical integrals appearing in the generated equations are nevertheless
all in folded form. For each unpinched massless banana sub-diagram, at most two kinds of
discrete-state representative can occur: the all-zero state,
\[
n_{e_1}=\cdots=n_{e_p}=0,
\]
and the odd-parity representative,
\[
n_{e_1}=1,\qquad n_{e_2}=\cdots=n_{e_p}=0.
\]
The fold therefore reduces the space of candidate integrals before reduction. The final
master-integral count is determined by the rank of the generated linear system.

\section{The \texttt{NuConvention} option}
\label{app:nu}

The initialization option \texttt{NuConvention} fixes the sign convention in which the package
writes its baseline formulas involving $\nu_e$. Under the default \texttt{"Positive"}, $\nu_e$ is
the Hankel order of \eqref{eq:bulkMode} exactly as supplied in the input; under
\texttt{"Negative"} every baseline formula is written with $-\nu_e$ instead. The choice is
resolved per massive line at initialization and frozen in the context; there is no per-call
override. The \texttt{"H"} preset and the Hankel functions themselves always consume the input
order $\nu_e$ and are insensitive to the choice; only the shapes of table~\ref{tab:nuconvention}
change. Writing $\bar\nu_e=\pm\nu_e$ for the two settings, the affected objects are as follows.

\begin{table}[!ht]
\centering
\small
\caption{Baseline shapes that depend on the \texttt{NuConvention} writing, with
$\bar\nu_e=+\nu_e$ under \texttt{"Positive"} and $\bar\nu_e=-\nu_e$ under \texttt{"Negative"}.}
\label{tab:nuconvention}
\renewcommand{\arraystretch}{1.15}
\begin{tabular}{p{5.6cm}p{4.2cm}}
\toprule
\textbf{Object} & \textbf{Shape} \\
\midrule
$P_{\mathrm h}(x)$, equation~\eqref{eq:functionSystem} & $(1-2\bar\nu_e)/x$ \\
equation-of-motion coefficient, \eqref{eq:eom} & $-(1-2\bar\nu_e)$ \\
Wronskian power of the \texttt{"h"} preset & $x^{\,2\bar\nu_e-1}$ \\
pinched-term zero point, \eqref{eq:pinched} & $a_{0u}+a_{0v}+2\bar\nu_e$ \\
residual sector zero power & $b_{S0,e}=b_{0e}-2\bar\nu_e$ \\
tree-propagator power in a pinched term & $2\bar\nu_e-1$ \\
small-time leading powers, states $0$ and $1$ & $\{2\bar\nu_e,0\}$, $\{2\bar\nu_e-1,1\}$ \\
naive tree DE, state-$1$ coefficient & $(1-2\bar\nu_e)/|k_e|$ \\
\bottomrule
\end{tabular}
\end{table}

The merged time power $a_u+a_v-1$, the residual momentum power
$b_{S,e}=b_e+1$ and all massless relations are convention independent. Since the two settings
differ by $\nu_e\to-\nu_e$ throughout, an input prepared for one writing produces, in the other
writing, the equations of the reflected mass parameter; comparisons against material frozen in
the opposite writing should declare the matching setting explicitly.

\section{Public functions}
\label{app:api}

Table~\ref{tab:api} lists the public functions. The dedicated options and specification keys of
each command are introduced together with that command in section~\ref{sec:using}, and
runtime option lists are available through \texttt{Options}.

\begin{table}[!ht]
\centering
\small
\caption{Public functions of the package.}
\label{tab:api}
\renewcommand{\arraystretch}{1.08}
\begin{tabular}{p{4.3cm}p{9.8cm}}
\toprule
\textbf{Function} & \textbf{Description} \\
\midrule
\texttt{DSInit} & initialize a case; topology, kinematic and time-domain audits; returns the
context. \\
\texttt{DSInfo} & brief or full information about the initialized context. \\
\texttt{DSKinematics} & two-stage kinematic proposal and validation. \\
\texttt{DSParameterNotation} & inspect the current parameter notation. \\
\texttt{DSRedefineParameters} & re-initialize the context with a new complete set of kinematic
rules. \\
\texttt{DSMessagesOn},\newline\texttt{DSMessagesOff},\newline\texttt{DSMessagesQ} & switch and query
the optional messages; fatal errors always stay on. \\
\texttt{DSSeeds} & complete set of IBP seed templates of a family. \\
\texttt{DSAllSeeds} & flattened seed template list. \\
\texttt{DSSeedGroups},\newline\texttt{DSSeedGroupMetadata} & grouping of the templates by sector, class and generator. \\
\texttt{DSMetaSeedRange} & per-group shift metadata for range planning. \\
\texttt{DSGenerateIBP} & sweep continuous indices; certified IBP relations. \\
\texttt{DSLinear} & backend-neutral linear system (\texttt{linearData}). \\
\texttt{DSReorderIntegrals} & re-index the full integral ordering. \\
\texttt{DSKiraPlan} & in-memory pre-reduction or formal reduction plan. \\
\texttt{DSKiraExport} & write Kira input directories; manifest with file identities. \\
\texttt{DSKiraImport} & read back external reduction results with identity checks. \\
\texttt{DSUserMI} & attach a user master-integral basis with rank checks. \\
\texttt{DSDE} & differential equations of the reduced basis. \\
\texttt{DSScaleCheck} & scaling relations of the family. \\
\texttt{DSTreeNaiveIBP},\newline\texttt{DSTreeNaiveDE} & naive tree-level IBP and DE checks. \\
\texttt{dtau}, \texttt{dqq}, \texttt{dqk}, \texttt{ds} & atomic generators and the full variable derivative. \\
\texttt{rep2innerform},\newline\texttt{rep2outform} & notation rewriters between user and internal forms. \\
\texttt{rep2Integrand} & expand $J$ into a formal integrand retaining named propagator blocks. \\
\texttt{symmetry} & apply the user and built-in tadpole symmetry relations of section~\ref{sec:seeds}. \\
\texttt{repSymmetry0} & the raw user symmetry relations of the topology input. \\
\texttt{DSPublicAPI} & the registry of public functions used by this table. \\
\bottomrule
\end{tabular}
\end{table}

\clearpage
\bibliographystyle{JHEP}
\bibliography{references}

\end{document}